\documentclass[conference,letterpaper]{IEEEtran}
\usepackage{etoolbox}
\newbool{showproofs}
\booltrue{showproofs}

\usepackage[utf8]{inputenc} 
\usepackage[T1]{fontenc}
\usepackage{url}
\usepackage{ifthen}
\usepackage{cite}
\usepackage[cmex10]{amsmath}
\usepackage{mleftright}   
\mleftright  

\usepackage{graphicx} 
\usepackage{booktabs}  
\usepackage{algpseudocode}

\usepackage{amsfonts,amssymb}
\usepackage{amsthm}
\usepackage{hhline}
\usepackage{multirow}
\usepackage{tikz} 
\usepackage{environ}

\usetikzlibrary{positioning}

\theoremstyle{definition}
\newtheorem{definition}{Definition}
\newtheorem{theorem}{Theorem}
\newtheorem*{theorem*}{Theorem}
\newtheorem{lemma}[theorem]{Lemma}
\newtheorem{example}{Example}
\newtheorem{claim}{Claim}

\newtheorem{observation}{Observation}

\newcommand{\bvec}[1]{\boldsymbol{#1}}
\newcommand{\Set}[1]{\mathcal{#1}}
\newcommand{\newdef}[1]{\textbf{\emph{#1}}}
\newcommand{\eqdef}{\triangleq}

\newcommand{\sv}[2]{_{[#1, #2]}}
\newcommand{\svtd}[2]{_{[#1; #2]}}

\newcommand{\w}[0]{\text{wt}}
\newcommand{\q}[0]{q}

\newcommand{\ldelta}{\ell, \delta}
\newcommand{\rv}{\text{read vector}} 
\newcommand{\Rv}[1]{R_{#1}} 
\newcommand{\Rcode}{read code} 
\newcommand{\Rchannel}{read channel} 

\newcommand{\capsymbol}{\mathsf{cap}}
\newcommand{\capacity}[1]{\capsymbol(#1)}
\newcommand{\capacityq}[2]{\capsymbol_{#1}(#2)}

\newcommand{\subvec}{sub-vector}

\newcommand{\ABC}{\Sigma_2}

\newcommand{\INT}{\mathbb{N}}

\newcommand{\AH}[0]{A_{\mathcal{H}^*}(\ldelta)}
\newcommand{\HG}[0]{\mathcal{H}(\ldelta)}
\newcommand{\HGs}[0]{\mathcal{H}^*(\ldelta)}
\newcommand{\G}[0]{\mathcal{G}(\ldelta)}
\newcommand{\V}[2]{V_{(#1, #2)}}
\newcommand{\cpol}[1]{p_{#1}(x)}

\newcommand{\di}[0]{\Delta}
\newcommand{\bi}[0]{\beta}
\newcommand{\psii}[0]{\psi}

\newcommand{\rmat}{\text{read matrix}}
\newcommand{\Rmat}[2]{R_{#1}^{#2}}

\begin{document}

\title{\textbf{The Capacity of the Weighted Read Channel}} 

\author{%
  \IEEEauthorblockN{\textbf{Omer Yerushalmi}, \textbf{Tuvi Etzion} and \textbf{Eitan Yaakobi}\vspace{-2ex}} \\
  \IEEEauthorblockA{Department of Computer Science, Technion — Israel Institute of Technology, Haifa, 3200003 Israel\\
  Email: \{omeryer, etzion, yaakobi\}@cs.technion.ac.il}\vspace{-5ex}
}

\maketitle

\begin{abstract}
    One of the primary sequencing methods gaining prominence in DNA storage is nanopore sequencing, attributed to various factors. 
    In this work, we consider a simplified model of the sequencer, characterized as a channel. This channel takes a sequence and processes it using a sliding window of length $\ell$, shifting the window by $\delta$ characters each time. The output of this channel, which we refer to as the \emph{\rv}, is a vector containing the sums of the entries in each of the windows. 
    The capacity of the channel is defined as the maximal information rate of the channel.
    Previous works have already revealed capacity values for certain parameters $\ell$ and $\delta$.
    In this work, we show that when $\delta < \ell < 2\delta$, the capacity value is given by $\frac{1}{\delta}\log_2 \frac{1}{2}(\ell+1+ \sqrt{(\ell+1)^2 - 4(\ell - \delta)(\ell-\delta +1)})$. Additionally, we construct an upper bound when $2\delta < \ell$.
    Finally, we extend the model to the two-dimensional case and present several results on its capacity.
\end{abstract}

\section{Introduction}
\label{sec:introduction}
DNA storage is an emerging technology driven by the increasing demand for data storage. Consequently, there has been significant progress in both synthesis and sequencing technologies \cite{synthesisTech1, synthesisTech2, synthesisTech3, MinION}.
One particular sequencing methodology, named the nanopore sequencer,
is mainly renowned for its support in long reads, low cost, and high probability \cite{nanopore1, nanopore2, nanopore3}.
The nanopore sequencing process proceeds as follows: when reading a DNA strand, its nucleotides traverse a pore sequentially. In this continuous process, a constant number of nucleotides, denoted by $\ell$, pass through the pore simultaneously each time.
The output of the reading process is determined by the values of each of the $\ell$ nucleotides.
While this sequencing technique is efficient in multiple aspects, it also presents some challenges. Primarily, the output of the reading process experiences inter-symbol interference (ISI) due to the dependence on the values of $\ell$ nucleotides simultaneously rather than just one.
Moreover, this process may frequently introduce random errors in the reading output, leading to occurrences
like duplications or deletions of certain nucleotides
Thus, various models have been proposed for the nanopore sequencer \cite{nanoporeModels, Abstract-Nanopor}. 

In this work, we focus on a specific model corresponding to the transverse-read model as outlined and studied in \cite{TR-CODE}, which was motivated by racetrack memories. This model has been proposed in \cite{Abstract-Nanopor}, and also studied in \cite{ECC-Nanopor}.
It constitutes a specific instance of the ISI channel, which is characterized by parameters $(\ell, \delta)$, and is denoted as $(\ell, \delta)$\emph{-weighted read channel}. From this point forward, we refer to it as the $(\ell, \delta)$-\emph{\Rchannel}.
This channel characterizes the reading operation of the sequencer as a sliding window of size $\ell$, shifting over the sequence in increments of $\delta$.
Thus, for a sequence $(x_1, x_2, \ldots, x_n)$, the initial read examines the first $\ell$ characters, $(x_1, x_2, \ldots, x_\ell)$, while the subsequent read occurs with a shift of $\delta$ characters, i.e., $(x_{\delta+1}, x_{\delta+2}, \ldots, x_{\delta+\ell})$. 
Each read produces a value corresponding to the values of the $\ell$ characters. In our case, for simplification, we concentrate on the cases where the emitting value is the sum of all these $\ell$ characters. 
It is apparent that distinct sequences may produce identical outputs under the $(\ell, \delta)$-\Rchannel. Therefore, our focus in this work is to study the capacity of the $(\ell, \delta)$-\Rchannel, denoted by $\capacity{\ell,\delta}$, and is defined as the logarithmic ratio between the number of outputs and inputs of the $(\ell, \delta)$-\Rchannel.

Several works have already studied this and similar models, focused on finding both the capacity \cite{TR-CODE, Abstract-Nanopor}, and error-correcting codes \cite{TR-CODE, Abstract-Nanopor, ECC-Nanopor} for the channel. 
In the subject of finding the capacity, both \cite{Abstract-Nanopor, TR-CODE} introduce algorithms for computing the capacity value with fixed parameters. In particular, \cite{Abstract-Nanopor} focused on the cases where $\delta = 1$, and the output of each read is a general function dependent on the $\ell$ characters. On the other hand, \cite{TR-CODE} focused on the case where the read function is the Hamming weight for any $\ell$ and $\delta$, which is the $(\ell, \delta)$-\Rchannel\ studied in this paper. More specifically,~\cite{TR-CODE} solved $\capacity{\ell,\delta}$ for the following cases: 1) $\ell\leq \delta$, 2) $\ell$ is a multiple of $\delta$, and 3) $\delta=2$ and $\ell=3,5,7$, using an algorithm that can be generalized for other values $\ell$. The main goal of this paper is to determine the capacity value $\capacity{\ell,\delta}$ for more parameters of $\ell$ and $\delta$.

The rest of this paper is organized as follows. In Section~\ref{sec:definitions}, we introduce the definitions describing the model. Section~\ref{sec:capacity-of-the-read-channel} is dedicated to solve the capacity $\capacity{\ell,\delta}$ for $\delta < \ell < 2\delta$, where it is shown that $\capacity{\ell,\delta}= \frac{1}{\delta}\log_2 \frac{1}{2}(\ell+1+ \sqrt{(\ell+1)^2 - 4(\ell - \delta)(\ell-\delta +1)})$. Furthermore, when $\ell>\delta$, in Section~\ref{sec:upper-bound-on-one-dim}, it is shown that $\capacity{\ell,\delta}\leq \frac{1}{\delta}\log_2 \frac{1}{2}(m+\sqrt{m^2+4m})$, where $m = (\ell \bmod \delta)((- \ell)\bmod \delta) + \delta$. Lastly, in Section~\ref{sec:additional-channels}, we extend the model to the two-dimensional case and present several results on the capacity as well. 
 
\section{Definitions and Preliminaries}
\label{sec:definitions}
Let $\ABC$ denote the binary alphabet $\{0, 1\}$.
For every vector $\bvec{x}\in \ABC^n$, we refer to its \subvec\ $(x_i, x_{i+1}, \ldots, x_{i+\ell-1})$, where $1\leq i\leq n-\ell$, as $\bvec{x}\sv{i}{\ell}$. The \emph{Hamming weight} of a vector $\bvec{x}$ is denoted by $\w(\bvec{x})$.
\begin{definition}
    The $(\ldelta)$-\newdef{\rv} of $\bvec{x}\in \ABC^n$ is denoted by, 
    \begin{align*}
        \Rv{\ldelta}(\bvec{x}) \eqdef (\w(\bvec{x}\sv{1}{\ell}), \w(\bvec{x}\sv{\delta+1}{\ell}), \ldots, \w(\bvec{x}\sv{t\cdot \delta +1}{\ell}))
    \end{align*}
    where $t = \frac{n-\ell}{\delta}$. For simplicity, we assume that $\delta | n-\ell$. 
    For each binary vector $\bvec{x}$, the $(\ldelta)$-\newdef{\Rchannel} produces the $(\ldelta)$-\rv\ of $\bvec{x}$.
\end{definition}
\begin{example}
    Let $\bvec{x}=(0, 0, 1, 0, 1, 0, 1, 1, 0, 0, 0, 0)$, the $(4, 2)$-\rv\ of $\bvec{x}$ is $\Rv{4, 2}(\bvec{x})=(1,2,3,2,0)$. 
    We can notice that there exist other vectors, such as $\bvec{y}=(0,0,0,1,0,1,1,0,0,0,0)$, $\bvec{y}\neq \bvec{x}$ that have the same \rv,
    i.e., $\Rv{4,2}(\bvec{y}) = \Rv{4,2}(\bvec{x})$.
\end{example}
Hence, a notable issue is that the \Rchannel\ might have the same output for multiple inputs. As a result, our main focus will be on assessing and describing this reduction. To achieve this, we will establish the following definitions:
\begin{definition}
    A code $\Set{C}\subseteq\ABC^n$ is called an $(\ldelta)$-\newdef{\Rcode} if for all distinct $\bvec{x}, \bvec{y}\in\Set{C}$ it holds that $\Rv{\ldelta}(\bvec{x}) \neq \Rv{\ldelta}(\bvec{y})$.
    The largest size of any $(\ldelta)$-\Rcode\ of length $n$ is denoted by $A(n, \ldelta)$.
    The \newdef{capacity} of the $(\ldelta)$-\Rchannel\ is given by:
    \begin{align*}
     \capacity{\ldelta} \eqdef \limsup_{n\rightarrow\infty}{\frac{\log_2A(n, \ldelta)}{n}}.
    \end{align*}
\end{definition}
A straightforward example of the $(\ldelta)$-\Rchannel, occurs when $\delta = 1$, as explored in \cite{TR-CODE}. In this case, all distinct vectors, $\bvec{x}, \bvec{y}\in \ABC^{n}$, have distinct $(\ell, 1)$-\rv, for every $\ell$ and $n$. Consequently, $A(n, \ell, 1) = 2^n$ and the capacity of the $(\ell, 1)$-\Rchannel\ is given by $\capacity{\ell, 1} = \limsup_{n\rightarrow\infty} \frac{1}{n}\log_2 2^{n}  = 1$. For the rest of the paper, it is assumed that $\delta>1$.

The binary model described here can be extended to the $\q$-ary case.
Within this model, the \emph{$L_1$ weight} is defined as the sum of all the entries in the vector and is also denoted by $\w(\bvec{x})$. The definitions remain the same as in the binary model, and here we refer to this channel as the $(\ldelta)_\q$-\Rchannel\ for the $\q$-ary alphabet, and we let $\capacityq{q}{\ldelta}$ denote its capacity. The capacity value in the $\q$-ary model can be directly deduced from the capacity in the binary case as is proved next in Theorem~\ref{theorem:binary-to-q-ary}. Consequently, our focus in this work is directed towards the binary case.

\begin{theorem} \label{theorem:binary-to-q-ary}
    Let $\ell, \delta$ be integers, for every integer $\q \geq 2$ it holds that,
    \vspace{-1.5ex}
    $$
        \capacityq{\q}{\ell, \delta} = \frac{\q-1}{\log_2 \q}\cdot \capacityq{2}{(\q-1) \cdot \ell, (\q-1) \cdot \delta}.
    $$
    \vspace{-3.5ex}  
\end{theorem}
\begin{IEEEproof}
    Let $\q\geq 2$ be an integer, and let $r = \q - 1$.
    The main goal of this proof is to demonstrate an injective mapping form every $(\ldelta)_\q$-\Rcode\ to a binary $(\ell', \delta')$-\Rcode\ and on the contrary, where $\delta' = r\cdot\delta$ and $\ell' = r\cdot\ell$, by converting between length-$r$ binary vectors and $\q$-ary characters.
    We begin by showing the first direction.
    Let $n\in \INT$, and let $\mathcal{C}$ be an $(\ldelta)_{\q}$-\Rcode\ of length $n$, such that $|\mathcal{C}| = A_{\q}(n, \ldelta)$. We define an injective mapping $\mu: \Sigma_{\q}^{n} \rightarrow \Sigma_{2}^{r\cdot n}$ from a vector in base $\q$ to a binary vector, such that every element $\alpha\in \Sigma_\q$ in the $\q$-ary vector is converted to the binary length-$r$ vector $1^{\alpha}0^{r-\alpha}$. i.e., for every $\bvec{x}\in \Sigma_\q^n$ and $0\leq i\leq n-1$, it holds that, $\mu(\bvec{x})\sv{r\cdot i+1}{r} = 1^{x_i}0^{r-x_i}$.
    By the definition of $\mu$, it holds that, for every $0\leq i\leq n-1$ $\w(x_{i+1}) = \w(\mu(\bvec{x})\sv{r\cdot i+1}{r})$, and therefore, for every $0\leq i\leq t$, $\w(\bvec{x}\sv{\delta\cdot i+1}{\ell}) = \w(\mu(\bvec{x})\sv{r\cdot\delta\cdot i+1}{r\cdot \ell})$. 
    Thus, $\Rv{\ldelta}(\bvec{x}) = \Rv{\ell', \delta'}(\mu(\bvec{x}))$. Let, $\mathcal{C}_\mu = \{\mu(\bvec{x}):\bvec{x}\in \mathcal{C}\}$. It holds that for every two distinct vectors $\bvec{x}, \bvec{y} \in \mathcal{C}$, $\Rv{\ldelta}(\bvec{x})\neq \Rv{\ldelta}(\bvec{y})$, and therefore, $\Rv{\ell', \delta'}(\mu(\bvec{x}))\neq \Rv{\ell', \delta'}(\mu(\bvec{y}))$. In conclusion, $\mathcal{C}_\mu$ is an $(\ell', \delta')$-\Rcode, and we have that $|\mathcal{C}_\mu| = |\mathcal{C}|$. Therefore, $A_{\q}(n, \ldelta) \leq A_2(r\cdot n, \ell', \delta')$.

    Now, we focus on the second direction. 
    Let $n\in \INT$, and let $\mathcal{C}$ be a binary $(\ell', \delta')$-\Rcode\ of length $r\cdot n$, such that $|\mathcal{C}| = A_{2}(r\cdot n, \ell', \delta')$. We define an injective mapping $\psi: \Sigma_{2}^{r\cdot n} \rightarrow \Sigma_{\q}^{n}$ from a binary vector to a vector in base $\q$, such that every \subvec\ $\bvec{y}$ starts at $r\cdot i+1$, $0\leq i \leq n-1$ of size $r$, is converted to the $\q$-ary element $\w(\bvec{y})\in \Sigma_\q$, i.e., for every $\bvec{x}\in \Sigma_2^{r\cdot n}$ and $0\leq i\leq n-1$, it holds that, $\mu(\bvec{x})_{i+1} = \w(\bvec{x}\sv{r\cdot i+1}{r})$.
    By the definition of $\psi$, it holds that, for every $0\leq i\leq n-1$ $\w(\psi(\bvec{x})_{i+1}) = \w(\bvec{x}\sv{r\cdot i+1}{r})$, and thus, by the same arguments as in the first direction, we have that $\mathcal{C}_\psi$ is an $(\ldelta)_{\q}$-\Rcode, $|\mathcal{C}_\psi| = |\mathcal{C}|$, and therefore $A_2(r\cdot n, \ell', \delta') \leq A_\q(n, \ldelta)$. In conclusion, $A_\q(n, \ldelta) = A_2(r\cdot n, \ell', \delta')$ for every integer $n$. And we get that,
    \begin{align*}
        \capacityq{\q}{\ldelta} = \frac{\log_\q A_\q(n, \ldelta)}{n}
        = \frac{\log_\q A_2(r\cdot n, \ell', \delta')}{n} \\
        = \frac{r}{\log_2 \q}\frac{\log_2 A_2( r\cdot n, \ell', \delta')}{r\cdot n} 
        = \frac{r}{\log_2 \q} \capacityq{2}{\ell', \delta'}.  
    \end{align*}
    
\end{IEEEproof}

This and similar models have been studied in several works, focusing on exploring the capacity \cite{TR-CODE, Abstract-Nanopor}, which focuses on finding expressions and bounds for the capacity.  Additionally, there is a concerted focus on finding error-correcting codes for the channel \cite{TR-CODE, Abstract-Nanopor, ECC-Nanopor} which are mainly focused on finding constructions and bounds on the size of the code in the cases where there is one deletion.
Our primary focus is on investigating the capacity of the $(\ldelta)$-\Rchannel\ across various parameters.
Multiple parameters of the $(\ldelta)$-\Rchannel\ have already been studied in \cite{TR-CODE}. 
First, explicit expressions and bounds for the capacity within the following parameters have been revealed.
\begin{theorem}[\hspace{0.01ex}\cite{TR-CODE}] 
    Let $\ell, \delta$ be positive integers,
    \begin{enumerate}
        \item For $\ell \leq \delta$, $\capacity{\ldelta} = \frac{1}{\delta}\log_2(\ell+1)$.
        \item If $\ell$ is a multiple of $\delta$, then $\capacity{\ldelta} = \frac{1}{\delta}\log_2(\delta+1)$.
        \item For $1 < \delta < \ell$, $\capacity{\ldelta} \geq \frac{1}{\delta}\log_2(\delta+1)$.
    \end{enumerate}
\end{theorem}
Second, the capacity value of the following cases, where $\delta = 2$ and $\delta < \ell$ have been calculated: 

\begin{scriptsize}
    \begin{center}
    \begin{tabular}{ |c|c|c|c|c|c|c| } 
     \hline
      & $\ell=3$ & $\ell=4$ & $\ell=5$ & $\ell=6$ & $\ell=7$ & $\ell=8$ \\ 
      \hline
     $\delta=2$ & 0.8857 & 0.7958 & 0.9258 & 0.7925 & 0.9361 & 0.7925  \\ 
     \hline
    \end{tabular}
    \end{center}
\end{scriptsize}

In this work, we present an explicit expression for $\capacity{\ldelta}$, where $\delta < \ell < 2\delta$. Additionally, we establish an upper bound on the capacity for the rest of the cases. Table~\ref{tab:my_label1} presents the current results on $\capacity{\ldelta}$, with entries in bold indicating new results derived from this work.

\vspace{-2ex}
\begin{table}[ht!]
    \centering
    \caption{}
    \vspace{-1ex}
    \begin{tabular}{|c|c||c|c|}
        \hline
        \multicolumn{2}{|c||}{$\ldelta>1$} & {$\capacity{\ldelta}$} \\
        \hline \hline 
        \multicolumn{2}{|c||}{$\ell\leq \delta$} & {$\capacity{\ldelta} = \frac{1}{\delta}\log_2(\ell+1)$} \\
        \hline 
        \multicolumn{2}{|c||}{$\delta < \ell < 2\delta$} & {$\capacity{\ldelta} = \boldsymbol{\frac{1}{\delta}\log_2 f_1(\ell, \delta)}$} \\ 
        \hline
        \multirow{2}{*}{$\ell\geq 2\delta$} & $\delta | \ell$ & {$\capacity{\ldelta} = \frac{1}{\delta}\log_2(\delta+1)$} \\
        \cline{2-3}
        & $\delta {\not|} \ell$ & $\frac{1}{\delta}\log_2(\delta+1) \leq \capacity{\ldelta} \leq  \boldsymbol{\frac{1}{\delta}\log_2 f_2(\ell, \delta)}$ \\
        \hline
    \end{tabular}
    \vspace{0.8ex}
    \\$f_1(\ell, \delta) = 0.5(\ell+1 + \sqrt{(\ell+1)^2 - 4(\ell-\delta)(\ell-\delta+1)})$, 
    \vspace{0.2ex}
    \\ $f_2(\ell, \delta) = 0.5(m-1 + \sqrt{(m-1)^2 - 4(m-1)})$,
    \vspace{0.2ex}
    \\and $m = (\ell \bmod \delta+ 1)(((-\ell) \bmod \delta+1))$.
    \label{tab:my_label1}
\end{table}

\section{The Capacity for $\delta < \ell < 2\delta$}
\label{sec:capacity-of-the-read-channel}
In this section, we study the value of $\capacity{\ldelta}$ when $\delta < \ell < 2\delta$.
First, we observe that the $(\ldelta)$-\Rchannel\ is a regular language and therefore can be recognized by a non-deterministic transition state diagram. In the next definition, we present such a diagram for any $\ell\geq\delta$. This diagram was proposed in~\cite{TR-CODE} and we present it here with an explanation of its correctness for the completeness of the results in the paper.
All edges in such a graph are labeled. Thus, we refer to any directed edge $(u, v)$ with a label $\alpha$ as $u \overset{\alpha}{\rightarrow} v$.
\begin{definition}
    The graph $\G$ is defined as follows.
    \begin{itemize}
        \item The nodes in $\G$ are the set of all vectors $\bvec{s}$ of length $\ell - \delta$, i.e., $V(\G) = \{\bvec{s}:\bvec{s}\in \ABC^{\ell-\delta}\}$.
        \item The set of directed labeled edges in $\G$ is defined as 
        \begin{align*}
            E(\G) \hspace{-0.7ex} = \hspace{-0.7ex} \{\bvec{x}\sv{1}{\ell-\delta} \hspace{-0.3ex} \overset{\alpha}{\rightarrow} \hspace{-0.3ex} \bvec{x}\sv{\delta+1}{\ell-\delta}\hspace{-0.3ex}:\hspace{-0.3ex}
            \bvec{x}\in \ABC^{\ell}, \alpha \hspace{-0.3ex} = \hspace{-0.3ex} \w(\bvec{x}) \hspace{-0.2ex} \}.
        \end{align*}
        That is, an edge between the nodes $\bvec{u}$ and $\bvec{v}$ with label $\alpha$ exists if there is a vector $\bvec{x}\in \ABC^{\ell}$ such that, $\bvec{u} = \bvec{x}\sv{1}{\ell-\delta}$, $\bvec{v} = \bvec{x}\sv{\delta+1}{\ell-\delta}$ and $\w(\bvec{x}) = \alpha$.
    \end{itemize}
\end{definition}
For every $n, t$, where $n = t\cdot \delta + \ell$,
every vector $\bvec{x}\in \Sigma_2^{n}$, has the following (unique) path $\bvec{x}\sv{1}{\ell - \delta} \overset{\alpha_0}{\rightarrow}\bvec{x}\sv{\delta + 1}{\ell - \delta}\overset{\alpha_1}{\rightarrow} \bvec{x}\sv{2\delta + 1}{\ell - \delta} \overset{\alpha_2}{\rightarrow} \cdots \overset{\alpha_t}{\rightarrow} \bvec{x}\sv{(t+1)\cdot \delta + 1}{\ell - \delta}$ in $\G$, such that $(\alpha_0,\alpha_1,\alpha_2,\ldots,\alpha_t)$ is the $(\ldelta)$-\rv\ of $\bvec{x}$. Therefore, there is a path in $\G$ for every \rv.
On the other hand, every path in $\G$ can correspond to more than one vector $\bvec{x}$, but to only one read vector. That is, all vectors $\bvec{x}$ that generate this path have the same read vector, and thus, $\G$ is a state diagram of the $(\ldelta)$-\Rchannel. 
Note that there might be two distinct vectors $\bvec{x}, \bvec{y}\in \ABC^\ell$, such that $\bvec{x}\sv{1}{\ell-\delta} = \bvec{y}\sv{1}{\ell-\delta} = \bvec{v}$ and $\bvec{x}\sv{\delta+1}{\ell-\delta} \neq \bvec{y}\sv{\delta+1}{\ell-\delta}$, with the same Hamming weight, denoted by $\alpha$. Therefore, the edges $\bvec{v} \overset{\alpha}{\rightarrow}\bvec{x}\sv{\delta+1}{\ell-\delta}$, and $\bvec{v} \overset{\alpha}{\rightarrow}\bvec{y}\sv{\delta+1}{\ell-\delta}$ exist in $\G$, and thus, $\G$ is not necessarily deterministic. 
In fact, for the case where $\delta < \ell \leq 2\delta$, $\G$ is a regular graph, and between any two nodes, there exists the same number of parallel edges, which is $2\delta - \ell +1$.

\begin{example} \label{example: non-det-grph}
    For $\delta = 3, \ell = 5$, the nodes set in $\G$, as shown in Fig.~\ref{fig:non-det-graph}, are the vectors $(0,0)$, $(0,1)$, $(1,0)$ and $(1, 1)$. 
    The vectors $(0, 0, 0, 0, 1)$ and $(0, 0, 1, 0, 0)$ are both in $\ABC^{\ell}$, and therefore, $(0, 0) \overset{1}{\rightarrow} (0, 1)$ and $(0, 0) \overset{1}{\rightarrow} (0, 0)$ are edges in $\G$. In conclusion, $\G$ is not deterministic.
    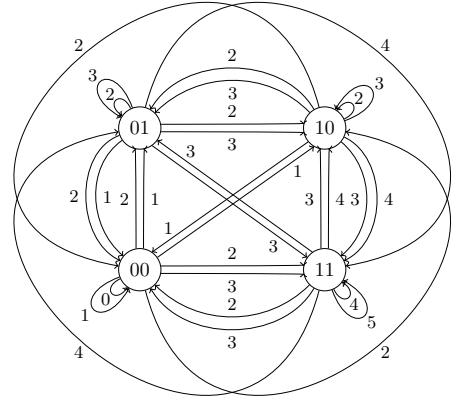
\begin{figure}[ht!]
        \centering
        \resizebox{\linewidth}{!}{
        \begin{tikzpicture}[node distance={25mm},  main/.style = {draw, circle}] 
        \useasboundingbox (-4,-2) rectangle (7.5, 4.5);
        \node[main] (0) {$00$};  
        \node[main] (1) [above of=0] {$01$}; 
        \node[main] (2) [right=25mm of 1] {$10$};
        \node[main] (3) [right=25mm of 0] {$11$};

        \draw[->] (0) to [out=205,in=235, looseness=5] node[midway,xshift=-3pt, yshift=-3pt]{\small$0$} (0);
        \draw[->] (0) to [out=200,in=240, looseness=10] node[midway, xshift=-4pt, yshift=-3pt]{\small$1$} (0);
        \draw[->] (0) to [out=10,in=170, looseness=0] node[midway, above]{\small$2$} (3);
        \draw[->] (0) to [out=350,in=190, looseness=0] node[midway, below]{\small$3$} (3);
        \draw[->] (0) to [out=100,in=260, looseness=0] node[midway, left]{\small$2$} (1);
        \draw[->] (0) to [out=80,in=280, looseness=0] node[midway, right]{\small$1$} (1);
        \draw[->] (0) to [out=35,in=240, looseness=0] node[pos=0.8, below]{\small$1$} (2);
        \draw[->, xshift=1.5cm] (0) to [out=285,in=350, looseness=3] node[midway, below]{\small$2$} (2);

        \draw[->] (1) to [out=115,in=145, looseness=5] node[midway,xshift=-3pt, yshift=3pt]{\small$2$} (1);
        \draw[->] (1) to [out=110,in=150, looseness=10] node[midway, xshift=-4pt, yshift=3pt]{\small$3$} (1);
        \draw[->] (1) to [out=10,in=170, looseness=0] node[midway, above]{\small$2$} (2);
        \draw[->] (1) to [out=350,in=190, looseness=0] node[midway, below]{\small$3$} (2);
        \draw[->] (1) to [out=215,in=140, looseness=1] node[midway, right]{\small$1$} (0);
        \draw[->] (1) to [out=200,in=155, looseness=1] node[midway, left]{\small$2$} (0);
        \draw[->] (1) to [out=305,in=145, looseness=0] node[pos=0.7, below]{\small$3$} (3);
        \draw[->] (1) to [out=75,in=10, looseness=3] node[midway, above]{\small$4$} (3);

        \draw[->] (2) to [out=25,in=55, looseness=5] node[midway,xshift=3pt, yshift=3pt]{\small$2$} (2);
        \draw[->] (2) to [out=20,in=60, looseness=10] node[midway, xshift=4pt, yshift=3pt]{\small$3$} (2);
        \draw[->] (2) to [out=120,in=60, looseness=1] node[midway, above]{\small$2$} (1);
        \draw[->] (2) to [out=134,in=45, looseness=1] node[midway, below]{\small$3$} (1);
        \draw[->] (2) to [out=320,in=40, looseness=1] node[midway, left]{\small$3$} (3);
        \draw[->] (2) to [out=335,in=25, looseness=1] node[midway, right]{\small$4$} (3);
        \draw[->] (2) to [out=220,in=55, looseness=0] node[pos=0.8, above]{\small$1$} (0);
        \draw[->] (2) to [out=105,in=170, looseness=3] node[midway, above]{\small$2$} (0);

        \draw[->] (3) to [out=295,in=325, looseness=5] node[midway,xshift=3pt, yshift=-3pt]{\small$4$} (3);
        \draw[->] (3) to [out=290,in=330, looseness=10] node[midway, xshift=4pt, yshift=-3pt]{\small$5$} (3);
        \draw[->] (3) to [out=225, in=315, looseness=1] node[midway, above]{\small$2$} (0);
        \draw[->] (3) to [out=240,in=300, looseness=1] node[midway, below]{\small$3$} (0);
        \draw[->] (3) to [out=100,in=260, looseness=0] node[midway, left]{\small$3$} (2);
        \draw[->] (3) to [out=80,in=280, looseness=0] node[midway, right]{\small$4$} (2);
        \draw[->] (3) to [out=125,in=325, looseness=0] node[pos=0.7, above]{\small$3$} (1);
        \draw[->] (3) to [out=255,in=190, looseness=3] node[midway, below]{\small$4$} (1);
        \end{tikzpicture} 
        }
        \caption{The graph $\G$ for $\ell = 5$ and $\delta=3$.}
        \label{fig:non-det-graph}
    \end{figure}
\end{example}
A common approach to deriving the capacity from a non-deterministic state diagram is to convert the graph to a deterministic one. Before doing so, let us introduce another useful definition.  
\begin{definition}
    For every subset $V$ of $V(\G)$, and a label $\alpha$.
    Let $\mathcal{E}(V, \alpha)\subseteq V(\G)$ be the set of all nodes with an incoming edge labeled $\alpha$ from any node in $V$, i.e., 
    \begin{align*}
        \mathcal{E}(V, \alpha) = \{\bvec{u}: \exists \bvec{v}\in V : \bvec{v} \overset{\alpha}{\rightarrow} \bvec{u} \in E(\G)\}.
    \end{align*}
\end{definition}
Next, we introduce a deterministic graph of $\G$.
\begin{definition}
    The graph $\HG$ is defined as follows.
    \begin{itemize}
        \item The nodes of the graph $\HG$ are the set of all subsets of nodes in $\G$, denoted as $\V{a}{b}$, where the Hamming weights of the nodes are between $a$ and $b$, $0\leq a \leq b\leq \ell- \delta$, i.e.,
        $$\hspace{-1ex}V(\HG) = \{ \V{a}{b}:
        0\leq a \leq b\leq \ell- \delta\},$$
        where,
        $\V{a}{b} = \{\bvec{s}\in \ABC^{\ell-\delta} : a \leq \w(\bvec{s})\leq b\}$.
        \item The set of directed labeled edges in $\HG$, denoted by $E(\HG)$, is defined as 
        \begin{align*}
            \{&\V{a_1}{b_1} \overset{\alpha}{\rightarrow} \V{a_2}{b_2} : \mathcal{E}(\V{a_1}{b_1} \hspace{-0.2ex}, \hspace{-0.2ex} \alpha) \subseteq \V{a_2}{b_2},\\
            &\exists \bvec{u}^1, \hspace{-0.5ex} \bvec{u}^2 \hspace{-0.5ex} \in \mathcal{E}(\V{a_1}{b_1}\hspace{-0.2ex}, \hspace{-0.2ex} \alpha), \w(\bvec{u}^1) \hspace{-0.4ex} = \hspace{-0.4ex} a_2, \w(\bvec{u}^2)\hspace{-0.4ex} =\hspace{-0.4ex} b_2\}.
        \end{align*}
        That is, an edge between the nodes $\V{a_1}{a_2}$ and $\V{a_2}{b_2}$ with label $\alpha$ exists if, 1) all nodes of $\G$ in $\mathcal{E}(\V{a_1}{a_2}, \alpha)$ belongs to $\V{a_2}{b_2}$, and 2) there are nodes $\bvec{u}^1, \bvec{u}^2 \in \mathcal{E}(\V{a_1}{a_2}, \alpha)$ such that $\w(\bvec{u}^1) = a_2$ and $\w(\bvec{u}^2) = b_2$.
    \end{itemize}
\end{definition}

\begin{example} \label{example:det-grph}
    For $\delta = 3, \ell = 5$, the graph $\HG$ of $\G$ from Example~\ref{example: non-det-grph}, is shown in Fig.~\ref{fig:det-graph-1}. The graph $\HG$ contains the edge $\V{0}{0}\overset{1}{\rightarrow}\V{0}{1}$, because, 1. $\mathcal{E}(\V{0}{0}, 1) = \{(0,0), (0,1), (1, 0)\} = \V{0}{1}$, and thus, $\mathcal{E}(\V{0}{0}, 1)\subseteq \V{0}{1}$. 2. Both $(0, 0) \overset{1}{\rightarrow}(0, 0)$ and $(0, 0) \overset{1}{\rightarrow}(0, 1)$ are edges in $\G$, while $\w(0, 0) = 0$ and $\w(0,1) = 1$.
    \begin{figure}[ht!]
    \centering
    \resizebox{\linewidth}{!}{
    \begin{tikzpicture}[node distance={25mm}, thick, main/.style = {draw, circle}] 
    \useasboundingbox (-7.5,-3.2) rectangle (7,3.2);
        \node[main] (0) {$\V{0}{2}$}; 
        \node[main] (1) [below left of=0] {$\V{0}{0}$}; 
        \node[main] (2) [below right of=0] {$\V{2}{2}$}; 
        \node[main] (3) [above left of=0] {$\V{0}{1}$}; 
        \node[main] (4) [above right of=0] {$\V{1}{2}$}; 
        \node[main] (5) [right=30mm of 0] {$\V{1}{1}$}; 
        \draw[->] (0) to [out=235,in=265, looseness=4] node[midway, xshift=-1pt, yshift=-4pt]{\small$2$} (0);
        \draw[->] (0) to [out=275,in=305, looseness=4] node[midway, xshift=1pt, yshift=-4pt]{\small$3$} (0);
        \draw[->] (1) to [out=150,in=210, looseness=4] node[midway, left]{$0$} (1);
        \draw[->] (3) to [out=150,in=210, looseness=4] node[midway, left]{$1$} (3); 
        \draw[->] (2) to [out=30,in=330, looseness=4] node[midway, right]{$5$} (2); 
        \draw[->] (4) to [out=30,in=330, looseness=4] node[midway, right]{$4$} (4); 
        \draw[->] (0) to [out=225,in=45, looseness=1] node[midway, above]{$0$} (1);
        \draw[->] (0) to [out=315,in=135, looseness=1] node[midway, above]{$5$} (2);
        \draw[->] (0) to [out=30,in=240, looseness=1] node[midway, below]{$4$} (4);
        \draw[->] (4) to [out=220,in=50, looseness=1] node[midway, above]{$3$} (0);
        \draw[->] (3) to [out=300,in=150, looseness=1] node[midway, below]{$2$} (0);
        \draw[->] (0) to [out=130,in=320, looseness=1] node[midway, above]{$1$} (3);
        \draw[->] (1) to [out=130,in=150, looseness=3] node[midway, above]{$2$} (4);
        \draw[->] (4) to [out=310,in=330, looseness=3] node[midway, below]{$1$} (1);
        \draw[->] (2) to [out=210,in=230, looseness=3] node[midway, below]{$3$} (3);
        \draw[->] (3) to [out=30,in=50, looseness=3] node[midway, above]{$4$} (2);
        \draw[->] (1) to [out=80,in=280, looseness=1] node[midway, right]{$1$} (3);
        \draw[->] (3) to [out=260,in=100, looseness=1] node[midway, left]{$0$} (1);
        \draw[->] (2) to [out=80,in=280, looseness=1] node[midway, right]{$4$} (4);
        \draw[->] (4) to [out=260,in=100, looseness=1] node[midway, left]{$5$} (2);
        \draw[->] (4) to [out=170,in=10, looseness=1] node[midway, above]{$2$} (3);
        \draw[->] (3) to [out=350,in=190, looseness=1] node[midway, below]{$3$} (4);
        \draw[->] (2) to [out=175,in=5, looseness=1] node[midway, above]{$2$} (1);
        \draw[->] (1) to [out=350,in=190, looseness=1] node[midway, below]{$3$} (2);

        \draw[->] (5) to [out=225,in=40, looseness=1] node[midway, below]{$4$} (2); 
        \draw[->] (5) to [out=270,in=280, looseness=1] node[midway, below]{$1$} (1); 
        \draw[->] (5) to [out=135,in=320, looseness=1] node[midway, above]{$3$} (4); 
        \draw[->] (5) to [out=90,in=80, looseness=1] node[midway, above]{$2$} (3); 
    \end{tikzpicture} 
    }
    \caption{The determinizing graph $\mathcal{H}(5, 3)$ of $\mathcal{G}(5,3)$ in Fig.~\ref{fig:non-det-graph}.}
    \label{fig:det-graph-1}
\end{figure}
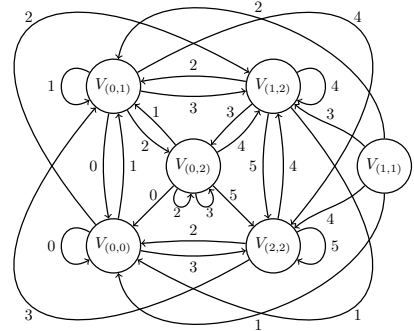
\end{example}
Next we prove that $\HG$ is a determinizing graph of $\G$.
\begin{claim}\label{cl:determinizing}
    The graph $\HG$ is a deterministic graph of $\G$.
\end{claim}
\begin{IEEEproof}
    First, we show that $\HG$ and $\G$ recognize the same language. 
    Let $\bvec{z} = z_1z_2\cdots z_n$ be a word generated by a path in $\G$, and let $\bvec{u}^0 \overset{z_1}{\rightarrow} \bvec{u}^1 \overset{z_2}{\rightarrow} \bvec{u}^2 \overset{z_3}{\rightarrow} \cdots \overset{z_n}{\rightarrow} \bvec{u}^{n}$ be such a path. 
    We show, by induction on $n$, the existence of a path $\V{a_0}{b_0} \overset{z_1}{\rightarrow} \V{a_1}{b_1} \overset{z_2}{\rightarrow} \cdots \overset{z_{n}}{\rightarrow} \V{a_{n}}{b_{n}}$ in $\HG$ where $\bvec{u}^n\in \V{a_{n}}{b_{n}}$.
    Base: $n=0$. Thus, $\bvec{z} = \varepsilon$, it holds that for every $\bvec{u}^0$ in $\G$, the zero-length path $\V{a}{a}$, where $a = \w(\bvec{u}^0)$, exists in $\HG$, and $\bvec{u}^0\in \V{a}{a}$.

    Step: Let $n \geq 1$, $\bvec{z} = z_1z_2\cdots z_n$ be a word generated by a path in $\G$, and let $\bvec{u}^0 \overset{z_1}{\rightarrow} \bvec{u}^1 \overset{z_2}{\rightarrow} \bvec{u}^2 \overset{z_3}{\rightarrow} \cdots \overset{z_n}{\rightarrow} \bvec{u}^{n}$ be such a path. By the induction assumption, we know that there exists a path $$\V{a_0}{b_0} \overset{z_1}{\rightarrow} \V{a_1}{b_1} \overset{z_2}{\rightarrow} \cdots \overset{z_{n-1}}{\rightarrow} \V{a_{n-1}}{b_{n-1}}$$ in $\HG$ where $\bvec{u}^{n-1} \in \V{a_{n-1}}{b_{n-1}}$. 
    Consider $a_n$ as the smallest and $b_n$ as the largest numbers such that $\bvec{s}_a\overset{z_{n}}{\rightarrow}\bvec{v}_a$ and $\bvec{s}_b\overset{z_{n}}{\rightarrow}\bvec{v}_b$ are edges in $\G$, where $\bvec{s}_a, \bvec{s}_b\in \V{a_{n-1}}{b_{n-1}}$ and $\w(\bvec{v}_a) = a_{n}$, $\w(\bvec{v}_b) = b_{n}$.There are such edges in $\G$, due to the existence of the edge $\bvec{u}^{n-1}\overset{z_{n}}{\rightarrow}\bvec{u}^{n}$ in $\G$. Therefore, the path $\V{a_0}{b_0} \overset{z_1}{\rightarrow} \V{a_1}{b_1} \overset{z_2}{\rightarrow} \cdots \overset{z_{n}}{\rightarrow} \V{a_{n}}{b_{n}}$ exists in $\HG$. In addition, $a_n\leq \w(\bvec{u}^{n})\leq b_n$ and therefore, $\bvec{u}^n\in \V{a_n}{b_n}$. Hence, we get that $\bvec{z}$ is generated by a path in $\HG$.
    
    To show the other direction, let $\bvec{z} = z_1z_2\ldots z_n$ be a word generated by a path in $\HG$, and let $\V{a_0}{b_0} \overset{z_1}{\rightarrow} \V{a_1}{b_1} \overset{z_2}{\rightarrow} \cdots \overset{z_{n}}{\rightarrow} \V{a_{n}}{b_{n}}$ be such a path. We show by induction on $n$ that for every $\bvec{u}^n\in \V{a_n}{b_n}$, there exists a path $\bvec{u}^0 \overset{z_1}{\rightarrow} \bvec{u}^1 \overset{z_2}{\rightarrow} \cdots \overset{z_n}{\rightarrow} \bvec{u}^{n}$ in $\G$. Base: $n=0$. The statement holds by definition.

    Step: Let $n \geq 1$, $\bvec{z} = z_1z_2\ldots z_n$ be a word generated by a path in $\HG$, and let $$\V{a_0}{b_0} \overset{z_1}{\rightarrow} \V{a_1}{b_1} \overset{z_2}{\rightarrow} \cdots \overset{z_{n}}{\rightarrow} \V{a_{n}}{b_{n}}$$ be such a path. By the induction assumption, it is known that for every $\bvec{u}^{n-1}\in \V{a_{n-1}}{a_{n-1}}$, there exists a path $\bvec{u}^0 \overset{z_1}{\rightarrow} \bvec{u}^1 \overset{z_2}{\rightarrow} \cdots \overset{z_{n-1}}{\rightarrow} \bvec{u}^{n-1}$ in $\G$.
    By the existence of the edge $\V{a_{n-1}}{b_{n-1}} \overset{z_n}{\rightarrow} \V{a_n}{b_n}$, we know there are $\bvec{u}^{n-1}, \bvec{v}^{n-1}\in\V{a_{n-1}}{b_{n-1}}$ and $\bvec{u}^n, \bvec{v}^n\in \V{a_n}{b_n}$ such that the edges $\bvec{u}^{n-1} \overset{z_n}{\rightarrow} \bvec{u}^{n}$, $\bvec{v}^{n-1} \overset{z_n}{\rightarrow} \bvec{v}^{n}$ exists in $\G$ and $\w(\bvec{u}^n) = a_n$, $\w(\bvec{v}^n) = b_n$. 
    Therefore, there are vectors $\bvec{x}, \bvec{y}\in \ABC^\ell$ such that, $\w(\bvec{x}) = \w(\bvec{y}) = z_n$, $a_{n-1} \leq \w(\bvec{x}\sv{1}{\ell-\delta}), \w(\bvec{y}\sv{1}{\ell-\delta}) \leq b_{n-1}$ and $\bvec{x}\sv{\delta+1}{\ell-\delta} = \bvec{u}^n$, $\bvec{y}\sv{\delta+1}{\ell-\delta} = \bvec{v}^n$. 
    Thus, for every $\bvec{s}^n\in \V{a_n}{b_n}$ there exists a vector $\bvec{x}' = $ with Hamming weight of $z_n$, such that $a_{n-1} \leq \w((\bvec{x}')\sv{1}{\ell-\delta})\leq b_{n-1}$ $(\bvec{x}')\sv{\delta+1}{\ell-\delta} = \bvec{s}^n$, i.e., the edge $(\bvec{x}')\sv{1}{\ell-\delta} \overset{z_n}{\rightarrow} \bvec{s}^n$ exists. Therefore, for every $\bvec{u}^n\in \V{a_n}{b_n}$, there exists a path $\bvec{u}^0 \overset{z_1}{\rightarrow} \bvec{u}^1 \overset{z_2}{\rightarrow} \cdots \overset{z_n}{\rightarrow} \bvec{u}^{n}$ in $\G$. In conclusion, $\bvec{z}$ is generated by a path in $\G$.
    
    Second, we establish that $\HG$ is deterministic. 
    Assume by contradiction that $\HG$ is not deterministic. 
    Therefore, there are two edges $\V{a}{b} \overset{\alpha}{\rightarrow} \V{a_1}{b_1}$ and $\V{a}{b} \overset{\alpha}{\rightarrow} \V{a_2}{b_2}$ in $\HG$, such that $\V{a_1}{b_1}\neq \V{a_2}{b_2}$. For the case where $b_1 \neq b_2$, without loss of generality (w.l.o.g), assume $b_1 < b_2$.
    As a result, there exists an edge $\bvec{s}^1 \overset{\alpha}{\rightarrow} \bvec{s}^2$ in $\G$, such that $a < \w(\bvec{s}^1) \leq b$ , $\w(\bvec{s}^2) = b_2$. This, however, leads to a contradiction with the fact that all the edges, with label $\alpha$, from nodes in $\V{a}{b}$ should go into nodes in $\V{a_1}{b_1}$. By symmetry, the same holds for the case where $a_1 \neq a_2$, and thus $\HG$ is deterministic.
\end{IEEEproof}

We have established that $\HG$ is a deterministic, finite state transition diagram of the regular language of the $(\ldelta)$-\Rchannel. To find the adjacency matrix of $\HG$, we start by determining the number of edges from $\V{a_1}{b_1}$ to $\V{a_2}{b_2}$.
\begin{claim} \label{claim:num-of-edges}
    \begin{enumerate}
    \item If $a_2 = 0$ and $b_2 = \ell - \delta$, then the number of edges is $\max\{0, 3\delta - 2\ell + b_1 - a_1 + 1\}$.
    \item If $a_2 = 0$ and $b_2 < \ell - \delta$, then the number of edges is $1$ when $b_2 \leq b_1 - a_1 + 2\delta - \ell$ and otherwise $0$.
    \item If $a_2 > 0$ and $b_2 = \ell - \delta$, then the number of edges is $1$ if $a_2 \geq 2\ell - 3\delta - b_1 + a_1$ and otherwise $0$.
    \item Otherwise, there is an edge only if $a_1 + b_2 = b_1 + a_2 +2\delta -\ell$.
\end{enumerate}
\end{claim}
\begin{IEEEproof}
    To determine the count of edges between two nodes, we look for the number of possible labels $\alpha$ for such edges.
    Let $0\leq \alpha\leq \ell-\delta$ where $\V{a_1}{b_1} \overset{\alpha}{\rightarrow} \V{a_2}{b_2}$ is an edge in $\HG$.
    If such an edge exists, then, there must be $\bvec{x}, \bvec{y}\in \ABC^\ell$ such that the values of both $\w(\bvec{x})$ and $\w(\bvec{y})$ are $\alpha$ and, $a_1\leq \w(\bvec{x}\sv{1}{\ell-\delta})$, $\w(\bvec{y}\sv{1}{\ell-\delta})\leq b_1$, $\w(\bvec{x}\sv{\delta+1}{\ell-\delta}) = b_2$, $\w(\bvec{y}\sv{\delta+1}{\ell-\delta}) = a_2$. 
    In addition, $0\leq \w(\bvec{x}\sv{\ell - \delta+1}{2\delta - \ell}), \w(\bvec{y}\sv{\ell - \delta+1}{2\delta - \ell}) \leq 2\delta - \ell$, and we get by adding the inequalities that,
    \begin{itemize}
        \item  $a_1 + b_2 \leq \w(\bvec{x}) \leq b_1 + b_2 + 2\delta - \ell$,
        \item $a_1 + a_2 \leq \w(\bvec{y}) \leq b_1 + a_2 + 2\delta - \ell$.
    \end{itemize}
    Therefore it holds that, $a_1+b_2 \leq \alpha \leq b_1 + a_2 + 2\delta - \ell$. In conclusion, 
    \begin{enumerate}
        \item If $a_2 = 0$ and $b_2 = \ell-\delta$, we have that
        $a_1 + \ell-\delta \leq \alpha \leq b_1 + 2\delta-\ell$. The number of options for such $\alpha$ is $\max\{0, 3\delta-2\ell+b_1-a_1 + 1\}$.
        \item If $a_2 = 0$ and $b_2 < \ell - \delta$, assume that there is an $\alpha > a_1 + b_2$.
        Thus, the vector $\bvec{x} = 1^{a_1}0^{t}1^{b_1+1}$, where $t = \ell - a_1 - b_1 - 1$, exists and holds that $\w(\bvec{x}) = \alpha$, $\bvec{x}\sv{1}{\ell-\delta}\in \V{a_1}{b_1}$, $\w(\bvec{x}\sv{\delta+1}{\ell-\delta}) = b_2+1$. This contradicts the fact that all edges in $\G$ from nodes in $\V{a_1}{b_1}$ go into nodes in $\V{a_2}{b_2}$. Therefore, $\alpha = a_1 + b_2$, and $a_1 + b_2 \leq b_1 + 2\delta - \ell$, and thus, if $a_1 + b_2 \leq b_1 + 2\delta - \ell$, one edge exists, and otherwise, there are zero edges.
        \item If $a_2 > 0$ and $b_2 = \ell - \delta$, assume that there is an $\alpha > b_1 + a_2 + 2\delta-\ell$.
        Thus, the vector $\bvec{x} = 1^{b_1}0^{t_1}1^{2\delta-\ell}0^{t_2}1^{a_1-1}$, where $t_1 = \ell-\delta-b_1$, and $t_2 = \ell - \delta - a_1 +1$, exists and holds that $\w(\bvec{x}) = \alpha$, $\bvec{x}\sv{1}{\ell-\delta}\in \V{a_1}{b_1}$, $\w(\bvec{x}\sv{\delta+1}{\ell-\delta}) = a_2-1$. 
        This contradicts the fact that all edges in $\G$ from nodes in $\V{a_1}{b_1}$ go into nodes in $\V{a_2}{b_2}$. Therefore, $\alpha = b_1 + a_2 + 2\delta-\ell$, and $a_1 + \ell - \delta \leq b_1 + a_2 + 2\ell - \delta$, and thus, if $a_1 - b_1 + 2\ell - 3\delta \leq a_2$, one edge exists, and otherwise, there are zero edges.
        \item Otherwise, $a_2 > 0$ and $b_2 < \ell -\delta$, from (2) and (3), we know that $\alpha = b_1 + a_2 + 2\delta - \ell = a_1 + b_2$, and thus, when $a_1 + b_2 = b_1 + a_2 +2\delta - \ell$, one edge exists, and otherwise, there are no edges.
    \end{enumerate}
\end{IEEEproof}

As observed in Example~\ref{example:det-grph}, the in or out degree of some nodes in $\HG$ might be zero. These nodes, as known, do not influence the value of the capacity and can thus be excluded from the graph.
We can see by Claim~\ref{claim:num-of-edges}, that the in and out degree of all the nodes $\V{a}{b}$ where $a = 0$ or $b = \ell - \delta$ is at least one. We denote $\Lambda_{\ldelta}$ to be the set of all such
nodes, i.e., $\Lambda_{\ldelta} \triangleq \{\V{0}{b}: 0\leq b \leq \ell-\delta\}\cup \{\V{a}{\ell-\delta}: 0\leq a\leq \ell-\delta\}$.
The in-degree of all other remaining nodes is at least one if  $b - a \geq 2\delta - \ell$, because in this case, there exist $a_1$ and $b_1$ such that $b_2 - a_2 = b_1 - a_1 + 2\delta - \ell$. All remaining nodes 
i.e., nodes which are not in $\Lambda_{\ldelta}$, and hold $b-a < 2\delta - \ell$, 
have an in-degree of zero. Let $\HGs$ denote the graph $\HG$ without those nodes.
\begin{example} \label{example:det-graph*}
    For $\delta = 3, \ell = 5$, the graph $\HGs$ of $\HG$ in Example~\ref{example:det-grph}, is the same graph as $\HG$ excluding the node $\V{1}{1}$ which its in-degree is zero.
    The graph is shown in Fig.~\ref{fig:det-graph}.
    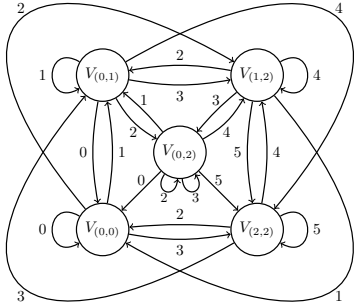
\begin{figure}[ht!]
    \centering
    \resizebox{\linewidth}{!}{
    \begin{tikzpicture}[node distance={25mm}, thick, main/.style = {draw, circle}] 
    \useasboundingbox (-8,-4) rectangle (7,3.2);
        \node[main] (0) {$\V{0}{2}$}; 
        \node[main] (1) [below left of=0] {$\V{0}{0}$}; 
        \node[main] (2) [below right of=0] {$\V{2}{2}$}; 
        \node[main] (3) [above left of=0] {$\V{0}{1}$}; 
        \node[main] (4) [above right of=0] {$\V{1}{2}$}; 
        \draw[->] (0) to [out=235,in=265, looseness=4] node[midway, xshift=-1pt, yshift=-4pt]{\small$2$} (0);
        \draw[->] (0) to [out=275,in=305, looseness=4] node[midway, xshift=1pt, yshift=-4pt]{\small$3$} (0);
        \draw[->] (1) to [out=150,in=210, looseness=4] node[midway, left]{$0$} (1);
        \draw[->] (3) to [out=150,in=210, looseness=4] node[midway, left]{$1$} (3); 
        \draw[->] (2) to [out=30,in=330, looseness=4] node[midway, right]{$5$} (2); 
        \draw[->] (4) to [out=30,in=330, looseness=4] node[midway, right]{$4$} (4); 
        \draw[->] (0) to [out=225,in=45, looseness=1] node[midway, above]{$0$} (1);
        \draw[->] (0) to [out=315,in=135, looseness=1] node[midway, above]{$5$} (2);
        \draw[->] (0) to [out=30,in=240, looseness=1] node[midway, below]{$4$} (4);
        \draw[->] (4) to [out=220,in=50, looseness=1] node[midway, above]{$3$} (0);
        \draw[->] (3) to [out=300,in=150, looseness=1] node[midway, below]{$2$} (0);
        \draw[->] (0) to [out=130,in=320, looseness=1] node[midway, above]{$1$} (3);
        \draw[->] (1) to [out=130,in=150, looseness=3] node[midway, above]{$2$} (4);
        \draw[->] (4) to [out=310,in=330, looseness=3] node[midway, below]{$1$} (1);
        \draw[->] (2) to [out=210,in=230, looseness=3] node[midway, below]{$3$} (3);
        \draw[->] (3) to [out=30,in=50, looseness=3] node[midway, above]{$4$} (2);
        \draw[->] (1) to [out=80,in=280, looseness=1] node[midway, right]{$1$} (3);
        \draw[->] (3) to [out=260,in=100, looseness=1] node[midway, left]{$0$} (1);
        \draw[->] (2) to [out=80,in=280, looseness=1] node[midway, right]{$4$} (4);
        \draw[->] (4) to [out=260,in=100, looseness=1] node[midway, left]{$5$} (2);
        \draw[->] (4) to [out=170,in=10, looseness=1] node[midway, above]{$2$} (3);
        \draw[->] (3) to [out=350,in=190, looseness=1] node[midway, below]{$3$} (4);
        \draw[->] (2) to [out=175,in=5, looseness=1] node[midway, above]{$2$} (1);
        \draw[->] (1) to [out=350,in=190, looseness=1] node[midway, below]{$3$} (2);
    \end{tikzpicture} 
    }
    \caption{The graph $\mathcal{H}^*(5, 3)$ of $\mathcal{H}(5,3)$ from Fig.~\ref{fig:det-graph-1}.}
    \label{fig:det-graph}
\end{figure}
\vspace{-1ex}
\end{example}
From now on, we focus on the graph $\HGs$. We observe from Claim~\ref{claim:num-of-edges} that the number of edges between $\V{a_1}{b_1}$ and $\V{a_2}{b_2}$ is exclusively determined by whether $\V{a_2}{b_2}$ is in $\Lambda_{\ldelta}$, as well as the values of $b_1 - a_1$ and $b_2 - a_2$. For every such node $\V{a}{b}$, we define its \emph{size} to be $b-a$.
For every $2\delta - \ell\leq d\leq \ell - \delta - 1$, the number of nodes which are not in $\Lambda_{\ldelta}$ and are of size $d$, denoted by $m_d$, is the number of options for nodes of size $d$ where $a\neq 0$ and $b\neq \ell-\delta$, i.e., $\ell -\delta - d - 1$.
Therefore, the total number of nodes in $\HGs$, denoted by $m$, is $|\Lambda_{\ldelta}| + \sum_{d = 2\delta-\ell}^{\ell-\delta-1}m_d = 1 + 2(\ell-\delta) + \binom{2\ell - 3\delta}{2}$.
Let $\AH \in \INT^{m\times m}$ be the adjacency matrix of $\HGs$ where its indices are ordered by the sizes of their nodes, while nodes with the same size are ordered lexicography. For shorthand, let $A_{i,j} \triangleq (\AH)_{i, j}$. We denote $t: \{1, \ldots, m\} \rightarrow \{\V{a}{b}:0\leq a \leq b \leq \ell-\delta\}$ to be a mapping between an index in the matrix to its node, and let $d: \{1, \ldots, m\} \rightarrow \{0, \ldots, \ell-\delta\}$ be a mapping between an index in the matrix to the size of its node. For example, $t(2) = \V{0}{\ell-\delta-1}$, and $d(2) = \ell - \delta - 1$. 
From Claim~\ref{claim:num-of-edges}, we get that for every $1\leq i \leq m$, $2 \leq j \leq m$,
\begin{itemize}
    \item $A_{i, 1} = \max\{0, 3\delta-2\ell + d(i) + 1\}$.
    \item If $t(j) \in \Lambda_{\ldelta}$, $d(j) \leq 2\delta - \ell + d(i)$, then $A_{i, j} = 1$.
    \item If $t(j)\notin \Lambda_{\ldelta}$, $d(j) = 2\delta - \ell + d(i)$, then $A_{i, j} = 1$.
    \item Otherwise, $A_{i, j} = 0$.
\end{itemize}

\begin{example}
    For $\delta=3$, $\ell = 5$, the adjacency matrix $\AH$ of $\HGs$ from Example~\ref{example:det-graph*}, is
    \begin{align*}
        A_{\mathcal{H}}(5,3)= 
        \begin{pmatrix}
            2 & 1 & 1 & 1 & 1 \\
            1 & 1 & 1 & 1 & 1 \\
            1 & 1 & 1 & 1 & 1 \\
            0 & 1 & 1 & 1 & 1 \\
            0 & 1 & 1 & 1 & 1 
        \end{pmatrix}
        \begin{matrix}
            &- &(\V{0}{2}) \\
            &- &(\V{0}{1}) \\
            &- &(\V{1}{2}) \\
            &- &(\V{0}{0}) \\
            &- &(\V{2}{2})
        \end{matrix}.
    \end{align*}
    To the right of the matrix, the values of $\V{a}{b}$ indicate the values of $t(i)$ for each index $i$.
\end{example}

\begin{claim} \label{claim:characteristic-polynomial}
    The characteristic polynomial of $\AH$, is 
    \begin{align*}
        \cpol{\ldelta} = (x^2 - (\ell + 1)x + (\ell - \delta)(\ell - \delta + 1))x^{m-2}.
    \end{align*}
\end{claim}

\begin{IEEEproof}
    To calculate $\cpol{\ldelta}$, we use the equivalent, $\cpol{\ldelta} = \det(xI - \AH)$. Thus, we now look at the determinant of $xI - \AH$. 
    After adding all columns to the first one and then subtracting the first row from all other rows, the result matrix satisfies the following properties:
    \begin{itemize}
        \item Each element in the first column, and the $i$-th row, $2\leq i\leq m$ is equal to $\sum_{j = 1}^{m} A_{1, j} - \sum_{j = 1}^{m} A_{i, j}$ and is denoted by $b_i$.
        \item Each element in the first row and the $j$-th column $2\leq j \leq m$ is equal to $-A_{1, j}$.
        \item All other elements are denoted by $a_{i, j}$, $2\leq i,j \leq m$, and are equal to $A_{1, j} - A_{i, j}$.
    \end{itemize}
    We start by explore the values of $a_{i, j} = A_{1, j} - A_{i, j}$ for every $2\leq j\leq m$.
    It holds that $A_{1, j} \neq A_{i, j}$ in two cases.
    First, if $t(j) \notin \Lambda_{\ldelta}$, and $d(j) = 2\delta - \ell +d(i)$, it holds that $A_{1, j} = 0, A_{i, j} = 1$.
    Second, if $t(j) \in \Lambda_{\ldelta}$, and $d(j) > 2\delta - \ell +d(i)$, then $A_{1, j} = 1, A_{i, j} = 0$.
    In all other cases, $A_{1, j} = A_{i, j}$. Note that, in particular, for every $i \leq j$ it holds that $d(j) < 2\delta -\ell +d(i)$, and therefore for every element above the main diagonal it holds that $A_{1, j} = A_{i, j}$.
    In addition, we look for the value of $\sum_{j = 1}^{m} A_{i, j}$ for every $1\leq i\leq m$. This sum is equal to the sum of $A_{i, 1}$ and the number of indices $2\leq j\leq m$, where $t(j)\in \Lambda_{\ldelta}$ and $d(j) \leq 2\delta-\ell + d(i)$, or where $t(j)\notin \Lambda_{\ldelta}$ and $d(j)=2\delta-\ell + d(i)$. 
    Hence, we get that $\sum_{j = 1}^{m} A_{i, j} = \delta + 1 + d(i)$, and in conclusion, $b_i = \sum_{j = 1}^{m} A_{1, j} - \sum_{j = 1}^{m} A_{i, j} = \ell - \delta - d(i)$.
    Therefore, we get that $\cpol{\ldelta}$ is calculated to be the determinant of the matrix

    \vspace{-2ex}
    \begin{small}
    \begin{align*}
    \begin{pmatrix}
        x \hspace{-0.6ex}-\hspace{-0.6ex} (\ell\hspace{-0.6ex}+\hspace{-0.6ex}1) & -A_{1,2} & -A_{1,3} & \cdots & -A_{1,m-1} & -A_{1,m} \\
        b_2 & x & 0 & \cdots & 0 & 0 \\
        b_3 & a_{3,2} & x & \cdots & 0 & 0 \\
        \vdots & \vdots & \vdots & \ddots & \vdots & \vdots\\ 
        b_{m-1} & a_{m-1,2} & a_{m-1,3} & \cdots & x & 0 \\
        b_{m} & a_{m,2} & a_{m,3} & \cdots & a_{m,m-1} & x
    \end{pmatrix}.
    \end{align*}
    \end{small}

    We now calculate the determinant by expanding along the first row. It holds that $-A_{1,j}$ equals $-1$ for every $j$ such that $t(j)\in \Lambda_{\ldelta}$, and otherwise it zero. Therefore
    \begin{align*}
        \cpol{\ldelta} = (x - (\ell + 1))x^{m-1} + \hspace{-1ex} \sum_{i: t(i) \in \Lambda_{\ldelta}} \hspace{-1ex}(-1)^{i}\det A(i),
    \end{align*}
    where $A(i)$, is in the form:
    \begin{align*}
        \begin{pmatrix}
        b_2 & x & \cdots & 0 & 0 & \cdots & 0 \\
        \vdots & \vdots & \ddots & \vdots & \vdots & \ddots & \vdots \\
        b_{i-1} & a_{i-1, 2} & \cdots & x & 0 & \cdots & 0 \\
        b_i & a_{i, 2} & \cdots & a_{i, i-1} & 0 & \cdots & 0 \\
        b_{i+1} & a_{i+1, 2} & \cdots & a_{i+1, i-1} & x & \cdots & 0 \\
        \vdots & \vdots & \ddots & \vdots & \vdots & \ddots & \vdots \\
        b_m & a_{m, 2} & \cdots & a_{m, i-1} & a_{m, i+1} & \cdots & x \\
    \end{pmatrix}.
    \end{align*}
    By relocating the row that starts with $b_{i}$ to become the first row through exchanges with all rows above it, we get that,
    \begin{align*}
        \cpol{\ldelta} = (x - (\ell + 1))x^{m-1} + \hspace{-1ex} \sum_{i: t(i) \in \Lambda_{\ldelta}} \hspace{-1ex}\det A'(i),
    \end{align*}
    where $A'(i)$, is in the form:
    \begin{align*}
        \begin{pmatrix}
        b_i & a_{i, 2} & \cdots & a_{i, i-1} & 0 & \cdots & 0 \\
        b_2 & x & \cdots & 0 & 0 & \cdots & 0 \\
        \vdots & \vdots & \ddots & \vdots & \vdots & \ddots & \vdots \\
        b_{i-1} & a_{i-1, 2} & \cdots & x & 0 & \cdots & 0 \\
        b_{i+1} & a_{i+1, 2} & \cdots & a_{i+1, i-1} & x & \cdots & 0 \\
        \vdots & \vdots & \ddots & \vdots & \vdots & \ddots & \vdots \\
        b_m & a_{m, 2} & \cdots & a_{m, i-1} & a_{m, i+1} & \cdots & x \\
    \end{pmatrix}.
    \end{align*}
    
    To calculate the determinant of $A(i)$, we define for every $n\in \INT$ and $1\leq i\leq n$ the following set of non-negative $n\times n$ matrices,
    \begin{align*}
        \mathbb{B}(i, n) = \left\{ B(i, n) =
        \begin{bmatrix}
            U(i) & \boldsymbol{0} \\
            V & L
        \end{bmatrix} \right\},
    \end{align*}
    where:
    \begin{enumerate}
        \item \begin{align*}
        U(i) =
        \begin{pmatrix}
            b_i & a_{i,2} & \ldots & a_{i, i-1} \\
            b_2 & x & \ldots & 0 \\
            \vdots & \vdots & \ddots & \vdots \\
            b_{i-1} & a_{i-1, 2} & \ldots & x \\
        \end{pmatrix},
    \end{align*}
    \item $V$ is any  matrix of size $(n-i+1)\times (i-1)$,
    \item $L$ is a lower triangular matrix, of size $(n-i+1)\times (n-i+1)$, where all the values in the diagonal are $x$.
    \item $\boldsymbol{0}$ is the all zero matrix of size, $(i-1)\times (n-i+1)$.
    \end{enumerate}
    We can see that $A(i)$ is in $\mathbb{B}(i, m-1)$.
    Now, we show by induction on $d(i)$, where $2\leq i\leq m$, and $0\leq d(i)\leq \ell - \delta - 1$, that for every integer $n$, the determinant of every matrix $B(i, n)$ in $\mathbb{B}(i, n)$ is equal to $b_ix^{n-1}$.
    
    Base: Let $i$ be an integer such that, $d(i) = \ell - \delta - 1$, and therefore, for every $j\geq 2$, $a_{i, j} = 0$ and thus, $B(i, n)$ is a lower triangular matrix, and the determinant is equal to $b_ix^{n-1}$.
    
    Step: Let $i$ be an integer such that, $2\delta-\ell \leq d(i) < \ell-\delta-1$. Let $s(i) = 2\delta - \ell + d(i)$, for every $2\leq j\leq n$ $a_{i, j} = 1$ if $\ell - \delta > d(j) > s(i)$ and $t(j) \in \Lambda_{\ldelta}$, let $\Lambda_{\ldelta}(1)$ be the set of indices satisfying these conditions.
    In addition, $a_{i, j} = -1$ if $d(j) = s(i)$ and $t(j)\notin \Lambda_{\ldelta}$, Let $\Lambda_{\ldelta}(2)$ be the set of indices satisfying these conditions. Therefore, $\det B(i, n)$, is in the form:
    \begin{align*}
        b_ix^{n-1}+\hspace{-1.2ex}\sum_{j\in \Lambda_{\ldelta}(1)} \hspace{-1ex} \det B(j, n-1)-\hspace{-1.2ex}\sum_{j\in \Lambda_{\ldelta}(2)} \hspace{-1ex} \det B(j, n-1).
    \end{align*} 
    For every $j$ in $\Lambda_{\ldelta}(1)\cup\Lambda_{\ldelta}(2)$, $d(j) \geq s(i)$, and thus, in particular $d(j) > d(i)$. Therefore, by the induction assumption, we have that
    \begin{align*}
        \det B(i, n) =  b_ix^{n-1} \hspace{-0.3ex}+\hspace{-1ex} \sum_{j\in \Lambda_{\ldelta}(1)} \hspace{-1ex} b_jx^{n-2} \hspace{-0.3ex}-\hspace{-1ex} \sum_{j\in \Lambda_{\ldelta}(2)} \hspace{-1ex} b_jx^{n-2}\\
        = b_ix^{m'-1} \hspace{-0.5ex}+\hspace{-0.5ex} 2\left(\sum_{d = s(i) + 1}^{\ell - \delta - 1} \hspace{-0.9ex} \ell\hspace{-0.4ex} - \hspace{-0.4ex} \delta \hspace{-0.4ex} - \hspace{-0.4ex} d\right) \hspace{-0.5ex}-\hspace{-0.5ex} m_{s(i)}(\ell - \delta - s(i)).
    \end{align*}
    It holds that $2\sum_{d = s(i) + 1}^{\ell - \delta - 1} \ell \hspace{-0.4ex} - \hspace{-0.4ex} \delta \hspace{-0.4ex} - \hspace{-0.4ex} d = m_{s(i)}(\ell - \delta - s(i))$, and therefore, $\det B(i, n) = b_ix^{n-1}$. In particular, we get that $\det A(i) = b_ix^{m'-1}$, and therefore,
    \begin{align*}
        \cpol{\ldelta} = (x^2 - (\ell +1)x)x^{m-2} + 2\sum_{i = 1}^{\ell - \delta} i \cdot x^{m-2}\\
        = (x^2 - (\ell + 1)x + (\ell - \delta)(\ell - \delta + 1))x^{m-2}.
    \end{align*}
\end{IEEEproof}

We are finally arrived at the point where we can establish the expression for the capacity.
\begin{theorem} \label{theorem:general-function-of-capcity-delta-2delta}
    For every integers $\ell$ and $\delta$, where $\delta < \ell < 2\delta$ the capacity of the $(\ldelta)$-\Rchannel\ is given by
    \begin{align*}
    \capacity{\ldelta} = \frac{\log_2 \frac{\ell+1 + \sqrt{(\ell+1)^2 - 4(\ell-\delta)(\ell-\delta+1)}}{2}}{\delta}.
    \end{align*}
\end{theorem}

\begin{IEEEproof}
By Perron-Frobenius Theorem \cite{IntroCoding} we know that,
\begin{align*}
\capacity{\ldelta} = \frac{\log_2 \lambda(\AH)}{\delta},
\end{align*}
and by Claim~\ref{claim:characteristic-polynomial} we get that
\begin{align*}
\lambda(\AH) \hspace{-0.5ex}=\hspace{-0.5ex} 
\frac{\ell\hspace{-0.5ex}+\hspace{-0.5ex}1 \hspace{-0.5ex}+\hspace{-0.5ex} \sqrt{(\ell+1)^2 - 4(\ell-\delta)(\ell-\delta+1)}}{2}.
\end{align*}
Therefore,
\begin{align*}
\capacity{\ldelta} = \frac{\log_2 \frac{\ell+1 + \sqrt{(\ell+1)^2 - 4(\ell-\delta)(\ell-\delta+1)}}{2}}{\delta}.
\end{align*}

\end{IEEEproof}

\section{An Upper Bound on the Capacity for $\ell \geq 2\delta $}
\label{sec:upper-bound-on-one-dim}
In this section, we propose an upper bound on the capacity for the rest of the cases where $\delta$ does not divide $\ell$, and $\ell \geq2\delta$. 
To accomplish this, we introduce a constrained system with a higher capacity than the one of the $(\ldelta)$-\Rchannel.
For the rest of this section, for every $\ell$ and $\delta$, such that $\delta {\not|} \ell$, $\ell \geq 2\delta$
and $n, t$ integers such that $n = \delta t + \ell$. Let $a = \lfloor\frac{\ell}{\delta}\rfloor$, $b = \ell \bmod \delta$, and $d = \delta - b$, note that $\ell = a\delta+b$. We introduce the following claims and definitions.

\begin{observation} \label{prop:p}
    For every $\bvec{v}\in \ABC^n$, and $0\leq i\leq t$, it holds that $\w(\bvec{v}\sv{\delta\cdot i+1}{\ell})$ equals to
    \vspace{-2ex}
    \begin{align*}
        \w(\bvec{v}\sv{\delta \cdot i + 1}{b}) \hspace{-0.4ex}+ \hspace{-0.5ex}
        \sum_{j = 1}^{a} \left(\w(\bvec{v}\sv{\delta \cdot (i + j) - d + 1}{d}) \hspace{-0.5ex}+\hspace{-0.5ex}\w(\bvec{v}\sv{\delta \cdot (i + j) + 1}{b})\right).
    \end{align*}
\end{observation}
\vspace{-1ex}
First, for shorthand, let $\di(i) \eqdef \delta i - d + 1$, for every $1\leq i\leq t + a$, and $\bi(i) \eqdef \delta i + 1$ for every $0\leq i\leq t + a$. Next,
as a result from Observation \ref{prop:p}, the $(\ldelta)$-\rv\ depends only on the weights of the sub-vectors $\bvec{v}\sv{\bi(i)}{b}$ and $\bvec{v}\sv{\di(j)}{d}$, where $0\leq i\leq t + a$, $1\leq j\leq t + a$. 
Therefore, the sequence of ones and zeros within these sub-vectors has no impact on the value of the read vector. 
Consequently, to establish an upper bound on the number of read vectors, we focus exclusively on vectors where all zeros appear before the ones in all sub-vectors $\bvec{v}\sv{\bi(i)}{b}$ and $\bvec{v}\sv{\di(i)}{d}$. Thus, we concentrate only on vectors in
$$\Pi_{\ldelta}^n \eqdef \Pi_b \times \Pi_d \times \cdots \times \Pi_b \subseteq \ABC^n,$$
where $\Pi_m \eqdef \{0^m1^{m-\alpha}:0\leq \alpha \leq m\}$. 
Note that $|\Pi_{\ldelta}^n| \leq A(n, \ldelta)$, and thus, $\Pi_{\ldelta}$ provides an upper bound on $\capacity{\ldelta}$.
To find a tighter bound, we introduce the following lemma.
\begin{lemma} \label{lemma:maintaining-wightes}
    For every $\bvec{v}, \bvec{u}\in \Pi_{\ldelta}^n$ and $1\leq i \leq t-1$, if
    \begin{small}
    \begin{align*}
        \w(\bvec{v}\sv{\di(i)}{\delta}) = \w(\bvec{u}\sv{\di(i)}{\delta}),& \quad
        \w(\bvec{v}\sv{\bi(i+a)}{\delta}) = \w(\bvec{u}\sv{\bi(i+a)}{\delta}), \\
        \w(\bvec{v}\sv{\bi(i)}{b}) + \w(\bvec{v}\sv{\bi(i+a)}{b}) &= \w(\bvec{u}\sv{\bi(i)}{b}) + \w(\bvec{u}\sv{\bi(i+a)}{b}),
    \end{align*} 
    \end{small}
    while all other sub-vectors are equal, then 
    $\Rv{\ldelta}(\bvec{v}) = \Rv{\ldelta}(\bvec{u})$.
\end{lemma}
\begin{IEEEproof}
Let $\bvec{v},\bvec{u}\in \Pi_{\ldelta}^n$ be such vectors.
First, we know that for every  integer $j$ such that $0\leq j \leq i - a - 1$ or $i+a + 1 \leq j \leq t$, $\bvec{v}\sv{\bi(j)}{\ell}=\bvec{u}\sv{\bi(j)}{\ell}$, and thus $\w(\bvec{u}\sv{\bi(j)}{\ell}) = \w(\bvec{v}\sv{\bi(j)}{\ell})$.
Next, for every $i - a \leq j \leq i - 1$, we know that $\w(\bvec{v}\sv{\di(i)}{\delta}) = \w(\bvec{u}\sv{\di(i)}{\delta})$ while the rest of the values in $\bvec{v}\sv{\bi(j)}{\ell}$ and $\bvec{u}\sv{\bi(j)}{\ell}$ are the same. Thus, $\w(\bvec{v}\sv{\bi(j)}{\ell}) = \w(\bvec{u}\sv{\bi(j)}{\ell})$.
By symmetry, the same holds for every $i + 1 \leq j \leq i + a$. 
Finally, in the case were $j = i$, we know that 
    $\w(\bvec{v}\sv{\bi(i)}{b}) + \w(\bvec{v}\sv{\bi(i+a)}{b}) = \w(\bvec{u}\sv{\bi(i)}{b}) + \w(\bvec{u}\sv{\bi(i+a)}{b})$
while the rest of the values in $\bvec{v}\sv{\bi(j)}{\ell}$ and $\bvec{u}\sv{\bi(j)}{\ell}$ are the same. Therefore, $\w(\bvec{v}\sv{\bi(j)}{\ell}) = \w(\bvec{u}\sv{\bi(j)}{\ell})$.
In conclusion, for every $0\leq j\leq t$, $\w(\bvec{v}\sv{\bi(j)}{\ell}) = \w(\bvec{u}\sv{\bi(j)}{\ell})$, and therefore $\Rv{\ldelta}(\bvec{v}) = \Rv{\ldelta}(\bvec{u})$.
\end{IEEEproof}
Thus, we introduce the following mapping designed to maintain the value of the $(\ell, \delta)$-\rv.
\begin{definition}
    Let $\phi^n_{\ldelta}: \Pi_{\ldelta}^{n} \rightarrow \Pi_{\ldelta}^{n}$ be a function that changes the value of any $\bvec{v}$ according to the following steps:
    For every $i = 1,\ldots, t-1$, if there exists $\bvec{u}\in \ABC^{\ell-2b}$ such that $\bvec{v}\sv{\di(i)}{\ell+2d}  = 1^d0^b\bvec{u}1^b0^d$, then $$\phi_{\ldelta}(\bvec{v})\sv{\di(i)}{\ell+2d} = 01^{d-1}0^{b-1}1\bvec{u}01^{b-1}0^{d-1}1.$$
\end{definition}
\begin{example}\label{example:phi}
    For $\ell = 8$, $\delta = 3$, and $n = 14$, we have that, $a = 2$, $b = 2$, $d = 1$. Let $\bvec{v} = (\boldsymbol{0},\boldsymbol{1},1,\boldsymbol{0},\boldsymbol{0},0,\boldsymbol{0},\boldsymbol{1},1,\boldsymbol{1},\boldsymbol{1},0,\boldsymbol{1},\boldsymbol{1})$, the bold values in $\bvec{v}$ are the sub-vectors $\bvec{v}\sv{\bi(i)}{b}$, while the rest are the sub-vectors $\bvec{v}\sv{\di(i)}{d}$. For $i = 1$, $\di(i) = 3$, $\bi(i) = 4$, $\bi(i+a) = 10$, and $\di(i+a+1) = 12$, and we have that $\bvec{v}\sv{\di(i)}{\ell+2d} = 100\bvec{u}110 = 1^d0^b\bvec{u}1^b0^d$. Thus, $\phi_{\ldelta}^n(\bvec{v})\sv{\di(i)}{\ell+2d} = 01^{d-1}0^{b-1}1\bvec{u}01^{b-1}0^{d-1}0$ and 
    $$\phi_{\ldelta}^n(\bvec{v}) = (\boldsymbol{0},\boldsymbol{1},0,\boldsymbol{0},\boldsymbol{1},0,\boldsymbol{0},\boldsymbol{1},1,\boldsymbol{0},\boldsymbol{1},1,\boldsymbol{1},\boldsymbol{1}).$$
    As we can see $\Rv{\ldelta}(\phi_{\ldelta}^n(\bvec{v})) = \Rv{\ldelta}(\bvec{v}) = (3, 4, 6)$.
\end{example}
Note that, every change of the function $\phi^n_{\ldelta}$, maintains the values of
$\w(\bvec{v}\sv{\di(i)}{\delta})$, $\w(\bvec{v}\sv{\bi(i+a)}{\delta})$, and $\w(\bvec{v}\sv{\bi(i)}{b}) + \w(\bvec{v}\sv{\bi(i+a)}{b})$, while all other sub-vectors are not changed. Thus, by Lemma~\ref{lemma:maintaining-wightes}, we have that $\Rv{\ldelta}(\bvec{v}) = \Rv{\ldelta}(\phi^n_{\ldelta}(\bvec{v}))$.
In addition, $\phi^n_{\ldelta}$ ensures that there are no sub-vectors $\bvec{v}\sv{\di(i)}{\ell+2d}=1^d0^b\bvec{u}1^b0^d$. Thus, we construct the following code from $\phi^n_{\ldelta}$.
\begin{figure}[ht!]
    \centering
    \newcommand{\drowsize}{0.8}
\begin{tikzpicture}
  \draw[thick] (0,0) rectangle (11*\drowsize,\drowsize);
  
  \foreach \x in {1,...,10}
    \draw (\x*\drowsize,0) rectangle ++(\drowsize,\drowsize);
  
  \node at (\drowsize*1-\drowsize/2,\drowsize/2) {$1$};
  \node at (\drowsize*2-\drowsize/2,\drowsize/2) {$0$};
  \node at (\drowsize*3-\drowsize/2,\drowsize/2) {$0$};
  \node at (\drowsize*8-\drowsize/2,\drowsize/2) {$1$};
  \node at (\drowsize*9-\drowsize/2,\drowsize/2) {$1$};
  \node at (\drowsize*10-\drowsize/2,\drowsize/2) {$0$};

  \foreach \i in {0,1,2,3}
    \draw[->,>=stealth,shorten >=2pt,shorten <=2pt] ({3*\drowsize*\i+\drowsize},\drowsize+0.35) -- ++(0,-0.35);
  
  \foreach \i in {1,2,3}
    \node at (3*\drowsize*\i+\drowsize+0.1,1.25) {\footnotesize$\delta\hspace{-0.2ex}(\hspace{-0.2ex}i\hspace{-0.7ex}+\hspace{-0.8ex}\i\hspace{-0.2ex})\hspace{-0.8ex}+\hspace{-0.8ex}1$};
    
  \node at (\drowsize,1.25) {\footnotesize$\delta \hspace{-0.2ex} i\hspace{-0.8ex}+\hspace{-0.8ex}1$};
  \node at (8.9*\drowsize,1.25) {\footnotesize$\delta \hspace{-0.3ex}i\hspace{-0.8ex}+\hspace{-0.8ex}\ell$};
  \draw[->,>=stealth,shorten >=2pt,shorten <=2pt] ({9*\drowsize},\drowsize+0.35) -- ++(0,-0.35);

  \node at (5.65*\drowsize,1.25) {\footnotesize$\delta\hspace{-0.2ex}(\hspace{-0.2ex}i\hspace{-0.8ex}-\hspace{-0.8ex}1\hspace{-0.2ex})\hspace{-0.2ex}\hspace{-0.8ex}+\hspace{-0.8ex}\ell$};
  \draw[->,>=stealth,shorten >=2pt,shorten <=2pt] ({6*\drowsize},\drowsize+0.35) -- ++(0,-0.35);
      
  \foreach \i in {0,1,2}
    \draw[dashed] (3*\i*\drowsize+\drowsize,-0.2) -- ++(2*\drowsize,0) node[midway, below] {$b$};
  
  \foreach \i in {0,1,2,3}
    \draw[thick] (3*\i*\drowsize,-0.2) -- ++(\drowsize,0) node[midway, below] {$d$};
    
\end{tikzpicture}
\begin{tikzpicture}
  \draw[thick] (0,0) rectangle (11*\drowsize,\drowsize);
  
  \foreach \x in {1,...,10}
    \draw (\x*\drowsize,0) rectangle ++(\drowsize,\drowsize);
  
  \node at (\drowsize*1-\drowsize/2,\drowsize/2) {$0$};
  \node at (\drowsize*2-\drowsize/2,\drowsize/2) {$0$};
  \node at (\drowsize*3-\drowsize/2,\drowsize/2) {$1$};
  \node at (\drowsize*8-\drowsize/2,\drowsize/2) {$0$};
  \node at (\drowsize*9-\drowsize/2,\drowsize/2) {$1$};
  \node at (\drowsize*10-\drowsize/2,\drowsize/2) {$1$};

  \foreach \i in {0,1,2,3}
    \draw[->,>=stealth,shorten >=2pt,shorten <=2pt] ({3*\drowsize*\i+\drowsize},\drowsize+0.35) -- ++(0,-0.35);
  
  \foreach \i in {1,2,3}
    \node at (3*\drowsize*\i+\drowsize+0.1,1.25) {\footnotesize$\delta\hspace{-0.2ex}(\hspace{-0.2ex}i\hspace{-0.7ex}+\hspace{-0.8ex}\i\hspace{-0.2ex})\hspace{-0.8ex}+\hspace{-0.8ex}1$};
    
  \node at (\drowsize,1.25) {\footnotesize$\delta \hspace{-0.2ex} i\hspace{-0.8ex}+\hspace{-0.8ex}1$};
  \node at (8.9*\drowsize,1.25) {\footnotesize$\delta \hspace{-0.3ex}i\hspace{-0.8ex}+\hspace{-0.8ex}\ell$};
  \draw[->,>=stealth,shorten >=2pt,shorten <=2pt] ({9*\drowsize},\drowsize+0.35) -- ++(0,-0.35);

  \node at (5.65*\drowsize,1.25) {\footnotesize$\delta\hspace{-0.2ex}(\hspace{-0.2ex}i\hspace{-0.8ex}-\hspace{-0.8ex}1\hspace{-0.2ex})\hspace{-0.2ex}\hspace{-0.8ex}+\hspace{-0.8ex}\ell$};
  \draw[->,>=stealth,shorten >=2pt,shorten <=2pt] ({6*\drowsize},\drowsize+0.35) -- ++(0,-0.35);
  
  \foreach \i in {0,1,2}
    \draw[dashed] (3*\i*\drowsize+\drowsize,-0.2) -- ++(2*\drowsize,0) node[midway, below] {$b$};
  
  \foreach \i in {0,1,2,3}
    \draw[thick] (3*\i*\drowsize,-0.2) -- ++(\drowsize,0) node[midway, below] {$d$};
    
\end{tikzpicture}
    \caption{Example for $\ell = 8$ and $\delta=3$. The second vector is the value gets by applying $\phi_{\ldelta}^n$ on the first vector.}
    \label{fig:phi-ld}
\end{figure}
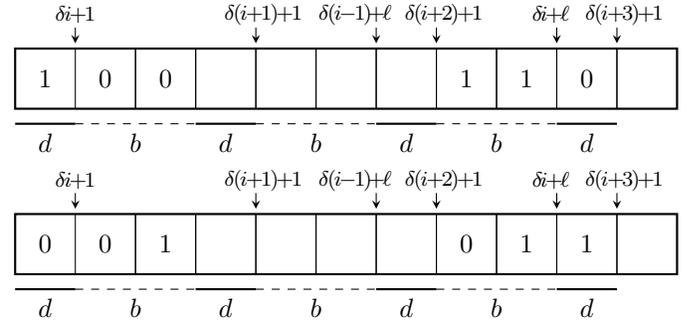

\begin{definition}
    The code $\mathcal{C}_{\ldelta}(n)$ is denoted to be the co-domain of $\phi_{\ldelta}^n(\bvec{v})$,
    \begin{align*}
    \mathcal{C}_{\ldelta}(n) \triangleq \{\phi_{\ldelta}^n(\bvec{v}) : \bvec{v}\in \Pi_{\ldelta}^n\}.
    \end{align*}
\end{definition}

Using the last observation on $\phi^n_{\ldelta}$, we can derive an upper bound on the value of $A(n, \ldelta)$.
\begin{lemma} \label{lemma:upper-bound-code}
    For every $n\in \INT$,
    $A(n, \ldelta) \leq |\mathcal{C}_{\ldelta}(n)|$.
\end{lemma}
\begin{IEEEproof}
    Let $C$ be a length $n$ $(\ldelta)$-\rv\ such that $|C| = A(n, \ldelta)$.
    We know that every iteration in $\phi_{\ldelta}^n$ meets the criteria of Lemma~\ref{lemma:maintaining-wightes} and therefore we have that the value of the read-vector does not change between any iteration, and therefore, we have that 
    \(
    \Rv{\ldelta}(\phi_{\ldelta}^n(\bvec{v})) = 
    \Rv{\ldelta}(\bvec{v})
    \).
    In addition, every vector in $\mathcal{C}$ has a different \rv\ and therefore, every vector in $\mathcal{C}$ has a different mapping in $\phi_{\ell,\delta}^n$. In conclusion, $\phi_{\ell,\delta}^n$ is an injective mapping from $\mathcal{C}$ to $\mathcal{C}_{\ldelta}(n)$. Hence, $|\mathcal{C}| \leq |\mathcal{C}_{\ldelta}(n)|$.   
\end{IEEEproof}
In Lemma~\ref{lemma:upper-bound-code}, we establish an upper bound for $A(n, \ldelta)$, however, determining the exact values of the bounds remains challenging. To address this, we introduce a more relaxed bound through the following constraint.

\begin{definition}
    Let $\mathcal{L}_{b, \delta}$ be the following constraint. First, the vector must be in $\Pi_{\ldelta}^n$, and second, every \subvec\ of length $2\delta$, that starts in an index of the form of $\di(i)$, is not in the form of $1^d0^b1^d0^b$, i.e., 
    $$
    \mathcal{L}_{b,\delta} = \{\bvec{v}:\forall i, \bvec{v}\sv{\di(i)}{2\delta}\neq 1^d0^b1^d0^b\}.
    $$
    Let $A(n, \mathcal{L}_{b, \delta})$, be the number of length-$n$ vectors that satisfy the constraint. The capacity of the constraint is denoted by $\capacity{\mathcal{L}_{b, \delta}}$, i.e., 
    \begin{align*}
        \capacity{\mathcal{L}_{\ell, \delta}} = \lim_{n\rightarrow\infty} \frac{\log_2 A(n, \mathcal{L}_{b, \delta})}{n}.
    \end{align*}
\end{definition}
Let $\mathcal{W}_{b, \delta}(n) \subseteq \Pi_{\ldelta}^{n}$ be a code that satisfies the $\mathcal{L}_{b, \delta}$ constraint such that $|\mathcal{W}_{b, \delta}(n)| = A(n, \mathcal{L}_{b, \delta})$. To show that $\capacity{\ldelta}\leq \capacity{\mathcal{L}_{b, \delta}}$, we first define the following function.
\begin{definition}
    Let $g:\mathcal{C}_{\ell, \delta}(n+a\delta)\rightarrow\{\bvec{u} \in \Pi_{\ldelta}^{n+a\delta}: \bvec{u}\sv{1}{n} \in \mathcal{W}_{b, \delta}(n)\}$ be a function, such that for each $\bvec{v}$, $g(\bvec{v})$ is the value of $\bvec{v}$ after the following steps:
    \\For $i = 1, 2, \ldots, t+a-1$,
    if $\bvec{v}\sv{\di(i)}{2\delta} = 1^d0^b1^d0^b$, then
    \begin{itemize}
        \item $\bvec{v}\sv{\bi(i+a)}{\delta} = 1^b0^d$,
        \item $\bvec{v}\sv{\di(i+1)}{d} = 0^{\gamma}1^{d-\gamma}$,
        \item $\bvec{v}\sv{\bi(i)}{b} = 0^{\alpha}1^{b-\alpha}$,
    \end{itemize}
    where $\alpha = \w(\bvec{v}\sv{\bi(i+a)}{b})$, and $\gamma = \w(\bvec{v}\sv{\di(i+a+1)}{d})$.
    This is, $g(\bvec{v})$, is the result of the following steps:
    \begin{algorithmic}[1]
        \State $\bvec{\nu} \gets \bvec{v}$
        \For{$i$ in $1, \ldots, t+a-1$}
                \State $\alpha = \w(\bvec{\nu}\sv{\bi(i+a)}{b})$
                \State $\gamma = \w(\bvec{\nu}\sv{\di(i+a+1)}{d})$
                \If{$\bvec{\nu}\sv{\di(i)}{2\delta} = 1^d0^b1^d0^b$}
                    \State $\bvec{\nu}\sv{\di(i+1)}{\delta} = 0^{\gamma}1^{d-\gamma}0^{\alpha}1^{b-\alpha}$
                    \State $\bvec{\nu}\sv{\bi(i+a)}{\delta} = 1^{b}0^{d}$
                \EndIf
        \EndFor
        \State \Return $\bvec\nu$ ($=g(\bvec{v})$)
    \end{algorithmic}
    The changes are well defined since $a \geq 2$, ensuring that there is no overlap between the changed sub-vectors.
    Note that, after every step, $\bvec{v}\sv{\di(i)}{2\delta}\neq 1^d0^b1^d0^b$, and if $\bvec{\nu}\sv{\di(i)}{2\delta}$ was equal to $1^d0^b1^d0^b$ then $\bvec{v}\sv{\di(i)}{\ell+2d} \neq 1^d0^b\bvec{u}1^b0^d$.
\end{definition}
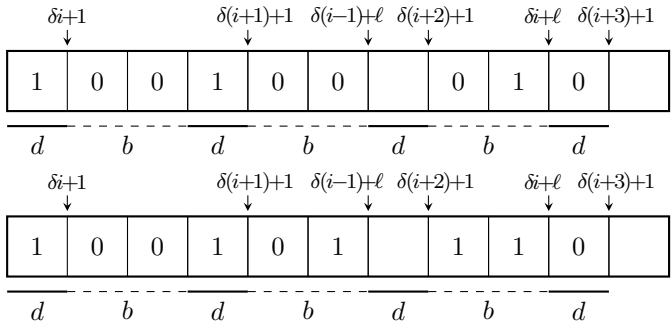
\begin{figure}[ht!]
    \centering
    \newcommand{\drowsize}{0.8}
\begin{tikzpicture}
  \draw[thick] (0,0) rectangle (11*\drowsize,\drowsize);
  
  \foreach \x in {1,...,10}
    \draw (\x*\drowsize,0) rectangle ++(\drowsize,\drowsize);
  
  \node at (\drowsize*1-\drowsize/2,\drowsize/2) {$1$};
  \node at (\drowsize*2-\drowsize/2,\drowsize/2) {$0$};
  \node at (\drowsize*3-\drowsize/2,\drowsize/2) {$0$};
  \node at (\drowsize*4-\drowsize/2,\drowsize/2) {$1$};
  \node at (\drowsize*5-\drowsize/2,\drowsize/2) {$0$};
  \node at (\drowsize*6-\drowsize/2,\drowsize/2) {$0$};
  \node at (\drowsize*8-\drowsize/2,\drowsize/2) {$0$};
  \node at (\drowsize*9-\drowsize/2,\drowsize/2) {$1$};
  \node at (\drowsize*10-\drowsize/2,\drowsize/2) {$0$};
  \foreach \i in {0,1,2,3}
    \draw[->,>=stealth,shorten >=2pt,shorten <=2pt] ({3*\drowsize*\i+\drowsize},\drowsize+0.35) -- ++(0,-0.35);
  
  \foreach \i in {1,2,3}
    \node at (3*\drowsize*\i+\drowsize+0.1,1.25) {\footnotesize$\delta\hspace{-0.2ex}(\hspace{-0.2ex}i\hspace{-0.7ex}+\hspace{-0.8ex}\i\hspace{-0.2ex})\hspace{-0.8ex}+\hspace{-0.8ex}1$};
    
  \node at (\drowsize,1.25) {\footnotesize$\delta \hspace{-0.2ex} i\hspace{-0.8ex}+\hspace{-0.8ex}1$};
  \node at (8.9*\drowsize,1.25) {\footnotesize$\delta \hspace{-0.3ex}i\hspace{-0.8ex}+\hspace{-0.8ex}\ell$};
  \draw[->,>=stealth,shorten >=2pt,shorten <=2pt] ({9*\drowsize},\drowsize+0.35) -- ++(0,-0.35);

  \node at (5.65*\drowsize,1.25) {\footnotesize$\delta\hspace{-0.2ex}(\hspace{-0.2ex}i\hspace{-0.8ex}-\hspace{-0.8ex}1\hspace{-0.2ex})\hspace{-0.2ex}\hspace{-0.8ex}+\hspace{-0.8ex}\ell$};
  \draw[->,>=stealth,shorten >=2pt,shorten <=2pt] ({6*\drowsize},\drowsize+0.35) -- ++(0,-0.35);
      
  \foreach \i in {0,1,2}
    \draw[dashed] (3*\i*\drowsize+\drowsize,-0.2) -- ++(2*\drowsize,0) node[midway, below] {$b$};
  
  \foreach \i in {0,1,2,3}
    \draw[thick] (3*\i*\drowsize,-0.2) -- ++(\drowsize,0) node[midway, below] {$d$};
    
\end{tikzpicture}
\begin{tikzpicture}
  \draw[thick] (0,0) rectangle (11*\drowsize,\drowsize);
  
  \foreach \x in {1,...,10}
    \draw (\x*\drowsize,0) rectangle ++(\drowsize,\drowsize);
  
  \node at (\drowsize*1-\drowsize/2,\drowsize/2) {$1$};
  \node at (\drowsize*2-\drowsize/2,\drowsize/2) {$0$};
  \node at (\drowsize*3-\drowsize/2,\drowsize/2) {$0$};
  \node at (\drowsize*4-\drowsize/2,\drowsize/2) {$1$};
  \node at (\drowsize*5-\drowsize/2,\drowsize/2) {$0$};
  \node at (\drowsize*6-\drowsize/2,\drowsize/2) {$1$};
  \node at (\drowsize*8-\drowsize/2,\drowsize/2) {$1$};
  \node at (\drowsize*9-\drowsize/2,\drowsize/2) {$1$};
  \node at (\drowsize*10-\drowsize/2,\drowsize/2) {$0$};
  \foreach \i in {0,1,2,3}
    \draw[->,>=stealth,shorten >=2pt,shorten <=2pt] ({3*\drowsize*\i+\drowsize},\drowsize+0.35) -- ++(0,-0.35);
  
  \foreach \i in {1,2,3}
    \node at (3*\drowsize*\i+\drowsize+0.1,1.25) {\footnotesize$\delta\hspace{-0.2ex}(\hspace{-0.2ex}i\hspace{-0.7ex}+\hspace{-0.8ex}\i\hspace{-0.2ex})\hspace{-0.8ex}+\hspace{-0.8ex}1$};
    
  \node at (\drowsize,1.25) {\footnotesize$\delta \hspace{-0.2ex} i\hspace{-0.8ex}+\hspace{-0.8ex}1$};
  \node at (8.9*\drowsize,1.25) {\footnotesize$\delta \hspace{-0.3ex}i\hspace{-0.8ex}+\hspace{-0.8ex}\ell$};
  \draw[->,>=stealth,shorten >=2pt,shorten <=2pt] ({9*\drowsize},\drowsize+0.35) -- ++(0,-0.35);

  \node at (5.65*\drowsize,1.25) {\footnotesize$\delta\hspace{-0.2ex}(\hspace{-0.2ex}i\hspace{-0.8ex}-\hspace{-0.8ex}1\hspace{-0.2ex})\hspace{-0.2ex}\hspace{-0.8ex}+\hspace{-0.8ex}\ell$};
  \draw[->,>=stealth,shorten >=2pt,shorten <=2pt] ({6*\drowsize},\drowsize+0.35) -- ++(0,-0.35);
  
  \foreach \i in {0,1,2}
    \draw[dashed] (3*\i*\drowsize+\drowsize,-0.2) -- ++(2*\drowsize,0) node[midway, below] {$b$};
  
  \foreach \i in {0,1,2,3}
    \draw[thick] (3*\i*\drowsize,-0.2) -- ++(\drowsize,0) node[midway, below] {$d$};
    
\end{tikzpicture}
    \caption{Example for the function $g$. The first vector represents the original vector, and the second vector is the result after applying the function $g$.
}
    \label{fig:g-ld}
\end{figure}
\begin{example}
    For the parameters as in Example~\ref{example:phi} and $n = 8$, let $\bvec{v}=(\boldsymbol{0},\boldsymbol{1},1,\boldsymbol{0},\boldsymbol{0},1,\boldsymbol{0},\boldsymbol{0},1,\boldsymbol{0},\boldsymbol{1},1,\boldsymbol{1},\boldsymbol{1})$. We can see that there is no $i$ such that $\bvec{v}\sv{\di(i)}{\ell+2d} = 1^d0^b\bvec{u}1^b0^d$, and thus $\bvec{v} \in \mathcal{C}_{\ldelta}(n+a\delta)$. We can see that for $i = 1$, $\bvec{v}\sv{\di(i)}{2\delta} = 1^d0^b1^d0^b$, and in addition, $\bvec{v}\sv{\bi(i+a)}{\delta} = 011$. Therefore,
    by applying $g$ on $\bvec{v}$, we get that after the first iteration $\bvec{v}\sv{\di(i+1)}{\delta} = 001$ and $\bvec{v}\sv{\bi(i+a)}{\delta} = 110$, i.e., 
    $$g(\bvec{v}) = (\boldsymbol{0},\boldsymbol{1},1,\boldsymbol{0},\boldsymbol{0},0,\boldsymbol{0},\boldsymbol{1},1,\boldsymbol{1},\boldsymbol{1},0,\boldsymbol{1},\boldsymbol{1}).$$
    We notice that now, $g(\bvec{v})\sv{\di(i)}{2\delta} \neq 1^d0^b1^d0^b$, while $g(\bvec{v})\sv{\di(i)}{\ell+2d} = 1^d0^b\bvec{u}1^b0^d$, and thus, $g(\bvec{v})$ is in $\mathcal{W}_{b, \delta}(n)$ and not in $C_{\ldelta}(n+a\delta)$.
\end{example}
Let $\bvec{\nu}(i)$ be the value of $\bvec{\nu}$ after the $i$-th iteration. Note that, $\bvec{\nu}(0) = \bvec{v}$ and $\bvec{\nu}(t+a-1) = g(\bvec{v})$. 
\begin{claim} \label{claim:mapping-ok}
    For every $\bvec{v}\in \mathcal{C}_{\ldelta}(n+a\delta)$, $g(\bvec{v})\sv{1}{n} \in \mathcal{W}_{b, \delta}(n)$.
\end{claim}
\begin{IEEEproof}
Let $\bvec{v}\in \mathcal{C}_{\ldelta}(n+a\delta)$.
We show by induction on $0\leq i\leq t+a-1$, that
\begin{enumerate}
    \item $\bvec{\nu}(i)\sv{\di(j)}{2\delta} \neq 1^d0^b1^d0^b$, for every $1\leq j\leq i$.
    \item $\bvec{\nu}(i)\sv{\di(i)}{\ell+2d} \neq 1^d0^b\bvec{u}1^b0^d$, for every $i < j \leq t+a-1$.
\end{enumerate}

Base: $i = 0$. The first statement holds by definition. In addition $\bvec{v}\in \mathcal{C}_{\delta}(n+a\delta)$ and $\bvec{\nu}(0) = \bvec{v}$, and therefore $\bvec{\nu}\sv{\di(j)}{\ell+2d} \neq 1^d0^b\bvec{u}1^b0^d$ for every $0\leq j\leq t+a-1$.

Step: Let $1\leq i\leq t+a-1$.
\begin{itemize}
    \item If $\bvec{\nu}(i-1)\sv{\di(i)}{2\delta}\neq1^d0^b1^d0^b$, then there was no change in the vector during the iteration, i.e., $\bvec{\nu}(i-1) = \bvec{\nu}(i)$. For the first statement, by the induction assumption, we know that $\bvec{\nu}(i-1)\sv{\di(j)}{2\delta}\neq 1^d0^b1^d0^b$, for every $1\leq j \leq i-1$. Therefore, $\bvec{\nu}(i)\sv{\di(j)}{2\delta} \neq 1^d0^b1^d0^b$, for every $1\leq j\leq i$.
    For the second statement, we know that $\bvec{\nu}(i-1)\sv{\di(j)}{\ell+2d} \neq 1^d0^b\bvec{u}1^b0^d$ for every $i-1< j \leq t+a-1$. Thus, because $\bvec{\nu}(i-1) = \bvec{\nu}(i)$, we get that $\bvec{\nu}(i)\sv{\di(j)}{\ell+2d} \neq 1^d0^b\bvec{u}1^b0^d$ for every $i < j \leq t+a-1$.
    \item Otherwise, $\bvec{\nu}(i-1)\sv{\di(i)}{2\delta} = 1^d0^b1^d0^b$.
    For the first statement, we know by the induction assumption that $\bvec{\nu}(i-1)\sv{\di(i)}{\ell+2d} \neq 1^d0^b\bvec{u}1^b0^d$, and therefore, $\bvec{\nu}(i-1)\sv{\bi(i+a)}{b} \neq 1^{b}$ or $\bvec{\nu}(i-1)\sv{\di(i+a+1)}{d} \neq 0^{d}$. Thus, after the iteration, $\bvec{\nu}(i)\sv{\di(i+1)}{d} \neq 1^{d}$ or $\bvec{\nu}(i)\sv{\bi(i+1)}{b} \neq 0^{b}$, and we get that $\bvec{\nu}(i)\sv{\di(i)}{2\delta}\neq 1^d0^b1^d0^b$. 
    In addition, for all $1\leq j < i$, there was not change in the sub-vector $\bvec{\nu}(i-1)\sv{\di(j)}{2\delta}$, i.e., $\bvec{\nu}(i)\sv{\di(j)}{2\delta} = \bvec{\nu}(i-1)\sv{\di(j)}{2\delta}$,
    and therefore, by the induction, for all $1\leq j \leq i$, $\bvec{\nu}(i)\sv{\di(j)}{2\delta} \neq 1^d0^b1^d0^b$.
    For the second statement, first we have that $\bvec{\nu}(i)\sv{\di(i+1)}{\delta} \neq 1^d0^b$, and therefore, $\bvec{\nu}(i)\sv{\di(i+1)}{\ell+2d} \neq 1^d0^b\bvec{u}1^b0^d$.
    Next, we have that for every $i+1 < j \leq t+a-1$, $\bvec{\nu}(i-1)\sv{\di(j)}{\ell+2d} \neq 1^d0^b\bvec{u}1^b0^d$,
    the changes in the sub-vector $\bvec{\nu}(i)\sv{\bi(i+a)}{\delta}$
    increase the number of zeros in the sub-vectors $\bvec{\nu}(i)\sv{\di(i+a+1)}{d}$, and the number of ones in the sub-vectors $\bvec{\nu}(i)\sv{\bi(i+a)}{b}$. Therefore, the changes do not create an index $i+1 < j$ such that $\bvec{\nu}(i)\sv{\di(j)}{\ell+2d} = 1^d0^b\bvec{u}1^b0^d$, and thus, for every $i < j \leq t+a-1$, we get that $\bvec{\nu}(i)\sv{\di(j)}{\ell+2d} \neq 1^d0^b\bvec{u}1^b0^d$. 
\end{itemize}
In conclusion, we get from the induction that for every $1\leq i \leq t+a-1$, $g(\bvec{v})\sv{\di(i)}{2\delta} = \bvec{\nu}(t+a-1)\sv{\di(i)}{2\delta} \neq 1^d0^b1^d0^b$ and therefore, $g(\bvec{v})\sv{1}{n}\in \mathcal{W}_{b, \delta}(n)$.
\end{IEEEproof}

 \begin{claim}\label{claim:g-injective}
    The function $g$ is injective.
\end{claim}
\begin{IEEEproof}
    Assume by contradiction that $g$ is not injective. Therefore, there are disjoint vectors $\bvec{v}^1, \bvec{v}^2\in \mathcal{C}_{\ldelta}(n+a\delta)$, such that $g(\bvec{v}^1) = g(\bvec{v}^2)$. Thus, we know that $\bvec{\nu}^1(0) \neq \bvec{\nu}^2(0)$ and that $\bvec{\nu}^1(t+a-1) = \bvec{\nu}^2(t+a-1)$.
    Let $0\leq i < t+a-1$, be the last iteration in the algorithm such that, $\bvec{\nu}^1(i) \neq \bvec{\nu}^2(i)$. Therefore, there must be a change in the value of at least one of the vectors. A change in both vectors is not possible, as they would remain unequal. Thus, there is a change in only one of the vectors, w.l.o.g., the change is in $\bvec{\nu}^1(i)$.
    We know that after the change, $\bvec{\nu}^1(i+1)\sv{\di(i+1)}{\ell+2d} = 1^d0^b\bvec{u}1^b0^d$, while by the induction in Claim~\ref{claim:mapping-ok}, we know that $\bvec{\nu}^2(i)\sv{\di(i+1)}{\ell+2d} \neq 1^d0^b\bvec{u}1^b0^d$. There was no change in $\bvec{\nu}^2(i)$, and therefore, $\bvec{\nu}^2(i+1)\sv{\di(i+1)}{\ell+2d} \neq 1^d0^b\bvec{u}1^b0^d$, and we got a contradiction to the fact that $\bvec{\nu}^1(i+1) = \bvec{\nu}^2(i+1)$.
\end{IEEEproof}
From Claims~\ref{claim:mapping-ok} and~\ref{claim:g-injective}, it holds that $|\{\bvec{v}\in \Pi_{\ldelta}^{n+a\delta}:\bvec{v}\sv{1}{n} \in \mathcal{W}_{b,\delta}(n)\}|\geq |\mathcal{C}_{\ldelta}(n+a\delta)|$, leading to the following upper bound.
\begin{theorem} \label{theorem:upper-bound-not-number}
For every $\ell, \delta$, such that $2\delta < \ell$ and $\delta{\not|} \ell$,
$\capacity{\mathcal{L}_{b, \delta}}$ is an upper bound of the capacity of the $(\ldelta)$-\Rchannel, i.e., 
\(
    \capacity{\ldelta} \leq \capacity{\mathcal{L}_{b, \delta}}
\), and
\begin{align*}
    \capacity{\mathcal{L}_{b, \delta}} = \frac{\log_2 \frac{m - 1 +\sqrt{(m - 1)^2 + 4(m-1)}}{2}}{\delta},
\end{align*}
where $m = (b+1)(d+1)$.
\end{theorem}

\begin{IEEEproof}
    First, by Claims~\ref{claim:mapping-ok} and~\ref{claim:g-injective}, we have that, $2^{a\delta}|\mathcal{W}_{b, \delta}(n)| \geq |\mathcal{C}_{\ell, \delta}(n+a\delta)|$.
    And by lemma~\ref{lemma:upper-bound-code} we got that,
    \begin{align*}
        \capacity{\ldelta} = & \lim_{n\rightarrow\infty} \frac{\log_2 A(n+a\delta, \ldelta)}{n+a\delta} \\
        \leq & \lim_{n\rightarrow\infty} \frac{\log_2 |\mathcal{C}_{\ldelta}(n+a\delta)|}{n+a\delta}\\ 
        \leq & \lim_{n\rightarrow\infty} \frac{\log_2 2^{a\delta}|\mathcal{W}_{b, \delta}(n)|}{n+a\delta} \\
        \leq & \lim_{n\rightarrow\infty} \frac{\log_2 A(n, \mathcal{L}_{b, \delta}) +a\delta}{n+a\delta} = \capacity{\mathcal{L}_{b, \delta}}.
    \end{align*}
    Second, for every $\delta > b > 0$, the deterministic state diagram of the language of $\mathcal{L}_{b, \delta}$ can be described as follows: The nodes in the state diagram are the vectors $\bvec{v}$ of length $\delta$, where $\bvec{v}\sv{1}{d} = 0^\alpha1^{d-\alpha}$ and $\bvec{v}\sv{d+1}{\delta} = 0^{\gamma}1^{b-\gamma}$, $0\leq \alpha \leq d$, $0\leq \gamma \leq b$, i.e., $$V = \{0^\alpha1^{d-\alpha}0^\gamma1^\gamma:0\leq \alpha \leq d, 0\leq \gamma \leq b\}.$$ 
    From every node $\bvec{v}$, there exists one edge to all nodes, excluding the node $1^d0^b$, which has no self-edge, i.e., 
    $$E = V\times V\setminus\{(1^d0^b, 1^d0^b)\}.$$ The label assigned to each node corresponds to the value of the next node.
    The number of nodes in the state diagram is $m = (b+1)(d+1)$. 
    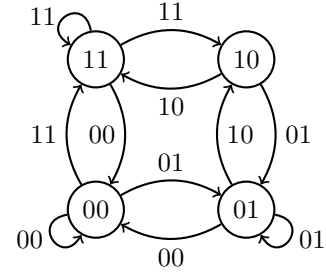
\begin{figure}[ht!]
        \centering
        \begin{tikzpicture}[node distance={20mm}, thick, main/.style = {draw, circle}] 
        \node[main] (0) {$00$}; 
        \node[main] (1) [right of=0] {$01$}; 
        \node[main] (2) [above of=1]{$10$}; 
        \node[main] (3) [above of=0] {$11$}; 
        \draw[->] (0) to [out=30,in=150, looseness=1] node[midway, above]{$01$} (1); 
        \draw[->] (1) to [out=210,in=330, looseness=1] node[midway, below]{$00$} (0); 
        \draw[->] (0) to [out=190,in=240, looseness=4] node[midway, left]{$00$} (0); 
        \draw[->] (1) to [out=350,in=300, looseness=4] node[midway, right]{$01$} (1); 
        \draw[->] (3) to [out=30,in=150, looseness=1] node[midway, above]{$11$} (2); 
        \draw[->] (2) to [out=210,in=330, looseness=1] node[midway, below]{$10$} (3); 
        \draw[->] (3) to [out=100,in=150, looseness=4] node[midway, left]{$11$} (3); 
        \draw[->] (0) to [out=120,in=240, looseness=1] node[midway, left]{$11$} (3); 
        \draw[->] (3) to [out=300,in=60, looseness=1] node[midway, left]{$00$} (0); 
        \draw[->] (1) to [out=120,in=240, looseness=1] node[midway, right]{$10$} (2); 
        \draw[->] (2) to [out=300,in=60, looseness=1] node[midway, right]{$01$} (1); 
        \end{tikzpicture} 
        \caption{Deterministic finite state diagram of $\mathcal{L}_{1, 2}$}
        \label{fig:non-det-graph-of-L}
    \end{figure}
    Thus, the adjacency matrix of the state diagram is all ones, except the cell $(i, i)$, where $i$ is the index of the node $1^d0^b$.
    Therefore, the maximal eigenvalue of the adjacency matrix is ${\lambda_{b, \delta} = \frac{m - 1 +\sqrt{(m - 1)^2 + 4(m-1)}}{2}}$, and 
    \begin{align*}
        \capacity{\mathcal{L}_{b, \delta}} = \frac{\log_2 \frac{m - 1 +\sqrt{(m - 1)^2 + 4(m-1)}}{2}}{\delta},
    \end{align*}
    where $m = (b+1)(d+1)$.
\end{IEEEproof}
The case where $\delta=2$ and $\ell\geq 3$ is odd has been studied in \cite{TR-CODE}. The question of whether a non-trivial upper bound, i.e., less than 1, exists for all these cases remained unsolved.
From Theorem~\ref{theorem:upper-bound-not-number} we can see that for every odd $\ell > 3$, 
\begin{align*}
    \capacity{\ell, 2} \leq \capacity{\mathcal{L}_{1,2}} = 0.9613.
\end{align*}
In addition, from Theorem~\ref{theorem:general-function-of-capcity-delta-2delta}  we have that $\capacity{3, 2} \leq 0.9613$, and therefore, we have that $0.9613$ is a non-trivial upper bound for all such cases.

\section{The Two-dimensional Weighted Channel}
\label{sec:additional-channels}
Coding for multiple dimensions, particularly two-dimensional storage systems, has gained significant attention in recent years due to its potential applications in diverse fields.
This is mainly due to the unique properties of the information which can be more accurately described in multiple dimensions.
Examples of such coding schemes were explored in \cite{ZCF}, \cite{ECMB}, \cite{TDCCT}, and \cite{BRBS}. 
Thus, we are interested in extending the \Rchannel\ to its two-dimensional version.
The two-dimensional \Rchannel\ 
is defined as follows.

For a matrix $B\in \Sigma_\q^{n_1\times n_2}$, let $B\svtd{k_1, \ell_1}{k_2, \ell_2}$ be the ${\ell_1\times \ell_2}$ sub-matrix of $B$ with entries between rows $k_1$ and ${k_1+\ell_1-1}$, and columns between $k_2$ and $k_2+\ell_2-1$.
The \emph{weight} of $B\svtd{k_1, \ell_1}{k_2, \ell_2}$ is denoted by $\w(B\svtd{k_1, \ell_1}{k_2, \ell_2})$ and is defined as the sum of entries in the window.
\begin{definition}
    The $((\ell_1, \ell_2), (\delta_1, \delta_2))$-\newdef{\rmat}, $\Rmat{\delta_1, \delta_2}{\ell_1, \ell_2}(B)$,
    of $B$ is a $(t_1+1)\times (t_2+1)$ matrix, where $t_1 = \frac{n_1-\ell_1}{\delta_1}$ and $t_2 = \frac{n_2-\ell_2}{\delta_2}$, and its $(i,j)$-th entry is defined by 
   $\w(B\svtd{k_1, \ell_1}{k_2, \ell_2})$,
    where $k_1 = \delta_1 i+1$, $ k_2 = \delta_2 j+1$, $0\leq i \leq t_1$,  and \break $0\leq j\leq t_2$.
\end{definition}
\begin{example}
    For $\q = 2$, $n_1=3$, $n_2=5$, ${\ell_1=2}$, ${\ell_2=3}$, ${\delta_1=1}$,  $\delta_2=2$, and
    \[
    B=
    \begin{pmatrix}
    1 & 0 & 1 & 1 & 1 \\
    0 & 0 & 1 & 0 & 1 \\
    0 & 1 & 0 & 1 & 1 \\
    \end{pmatrix}.
    \]
    The $((\delta_1, \delta_2), (\ell_1, \ell_2))$-\rmat\ of $B$ is,
    \[
    \Rmat{\delta_1, \delta_2}{\ell_1, \ell_2}(B) \hspace{-0.5ex}=\hspace{-0.5ex}
    \begin{pmatrix}
    \w
    \begin{pmatrix}
    1 &\hspace{-0.5ex} 0 &\hspace{-0.5ex} 1 \\
    0 &\hspace{-0.5ex} 0 &\hspace{-0.5ex} 1 \\
    \end{pmatrix} 
    &\hspace{-0.5ex}
    \w
    \begin{pmatrix}
    1 &\hspace{-0.5ex} 1 &\hspace{-0.5ex} 1 \\
    1 &\hspace{-0.5ex} 0 &\hspace{-0.5ex} 1 \\
    \end{pmatrix} \\\\
    \w
    \begin{pmatrix}
    0 &\hspace{-0.5ex} 0 &\hspace{-0.5ex} 1 \\
    0 &\hspace{-0.5ex} 1 &\hspace{-0.5ex} 0 \\
    \end{pmatrix} 
    &\hspace{-0.5ex}
    \w
    \begin{pmatrix}
    1 &\hspace{-0.5ex} 0 &\hspace{-0.5ex} 1 \\
    0 &\hspace{-0.5ex} 1 &\hspace{-0.5ex} 1 \\
    \end{pmatrix}
    \end{pmatrix}
    \hspace{-0.5ex}=\hspace{-0.5ex}
    \begin{pmatrix}
    3 & 5 \\
    2 & 4 \\
    \end{pmatrix}.
    \]
    Notice that there exist other matrices, e.g.,
    \[
    D=
    \begin{pmatrix}
    0 & 1 & 1 & 1 & 1 \\
    0 & 0 & 1 & 0 & 1 \\
    0 & 1 & 0 & 1 & 1 \\
    \end{pmatrix}
    \]
    different from $B$, such that $\Rmat{\delta_1, \delta_2}{\ell_1, \ell_2}(B)=\Rmat{\delta_1, \delta_2}{\ell_1, \ell_2}(D)$.
\end{example}
\begin{definition}
    A code $\mathbb{C}\subseteq\Sigma_\q^{n_1\times n_2}$ is called an $((\ell_1, \ell_2), (\delta_1, \delta_2))_{\q}$-\newdef{\Rcode} if for all
    distinct ${B, D\in\mathbb{C}}$ we have that
    $\Rmat{\delta_1, \delta_2}{\ell_1, \ell_2}(B) \neq \Rmat{\delta_1, \delta_2}{\ell_1, \ell_2}(D)$.
    The largest size of any $n_1\times n_2$ $((\ell_1, \ell_2), (\delta_1, \delta_2))_\q$-\Rcode\ is denoted by $A_\q(n_1,n_2,(\ell_1, \ell_2), (\delta_1, \delta_2))$.
    The \newdef{capacity} of the $((\ell_1, \ell_2), (\delta_1, \delta_2))_{\q}$-\Rchannel\ denoted by $\capacityq{\q}{(\ell_1, \ell_2), (\delta_1, \delta_2)}$ is defined by:
    \begin{align*}
     \limsup_{n, m\rightarrow\infty}{\frac{\log_\q A_{\q}(n_1,n_2,(\ell_1, \ell_2), (\delta_1, \delta_2))}{n_1\cdot n_2}}.
    \end{align*}
\end{definition}

\begin{theorem}
\label{theorem:binary-to-q-ary-two-d}
    For positive integers $\delta_1$, $\delta_2$, $\ell_1$, $\ell_2$, we have that for every two integers $q_1,q_2 \geq 1$ such that $q = q_1q_2+1$, the capacity of the $((\ell_1, \ell_2), (\delta_1, \delta_2))_{\q}$-\Rchannel, is equals to
    \begin{align*}
        \frac{\q_1\cdot \q_2}{\log_2 (\q_1\cdot \q_2+1)}\hspace{-0.3ex}\cdot\hspace{-0.3ex} \capacityq{2}{(\q_1 \cdot \ell_1, \q_2\cdot \ell_2), (\q_1\cdot \delta_1, \q_2\cdot \delta_2)}.
    \end{align*}
\end{theorem}
\begin{IEEEproof}
    Let $n_1$, $n_2$, $t_1$, $t_2$ be integers such that ${n_i = t_i\cdot \delta_i + \ell_i}$, and let $\ell_i' = \q_i'\ell_i$, $\delta_i' = \q_i'\delta_i$, $i = 1, 2$.
    We define an injective mapping
    $$
    \mu : \Sigma_{\q}^{n_1\times n_2}\rightarrow \Sigma_2^{(\q_1\cdot n_1)\times (\q_2\cdot n_2)}.
    $$
    Each element $\alpha \in \Sigma_{\q}$ of a matrix $B$ is mapped by $\mu$ into a binary
    $\q_1\times \q_2$ matrix by applying row by row folding on the vector $0^{\q_1\cdot \q_2-\alpha}1^{\alpha}$.
    The obtained $q_1 \times q_2$ matrices from each element of $\Sigma_q$ from $B$
    are concatenated in the same way as the elements of $B$ are concatenated in $B$.
    Note that, by this definition of $\mu$, for every $B\in \Sigma_{\q}^{n_1\times n_2}$, $0\leq i\leq t_1$, and $0\leq j\leq t_2$, we have that $\w(B\svtd{\delta_1 i +1, \ell_1}{\delta_2+1, \ell_2}) = \w(\mu(B)\svtd{\delta_1' i +1, \ell_1'}{\delta_2'+1, \ell_2'})$.
    For every $((\ell_1, \ell_2), (\delta_1, \delta_2))_\q$-\Rcode\ $\mathbb{C}$, let $\mathbb{C}_{\mu}$ be the code created by applying $\mu$ on the all matrices of $\mathbb{C}$, i.e., $\mathbb{C}_\mu = \{\mu(B) : B\in \mathbb{C}\}$. The function $\mu$ is injective and therefore, $|\mathbb{C}_\mu| = |\mathbb{C}|$.
    For every $B, D\in \mathbb{C}$, $\Rmat{\delta_1, \delta_2}{\ell_1, \ell_2}(B)\neq \Rmat{\delta_1, \delta_2}{\ell_1, \ell_2}(D)$, and thus, $\Rmat{\delta_1', \delta_2'}{\ell_1', \ell_2'}(\mu(B))\neq\Rmat{\delta_1', \delta_2'}{\ell_1', \ell_2'}(\mu(D))$. In conclusion, $\mathbb{C}_{\mu}$ is
    an $(\ell_1', \ell_2'), (\delta_1', \delta_2')$-\Rcode, and therefore, 
    \begin{align}
        A_{\q}(n_1, n_2, &(\ell_1, \ell_2), (\delta_1, \delta_2)) \notag\\
        &\leq A(\q_1\cdot n_1, \q_2\cdot n_2, (\ell_1', \ell_2'), (\delta_1', \delta_2')).\label{eqn:1}
    \end{align}
    Now, we define a mapping
    $$\lambda:\Sigma_2^{(\q_1\cdot n_1)\times (\q_2\cdot n_2)}\rightarrow\Sigma_{\q}^{n_1\times n_2},$$
    such that every $\q_1\times \q_2$ sub-matrix $B$ is mapped into an element $\beta = \w(B)\in \Sigma_\q$, i.e., for every matrix $B\in \Sigma_2^{(\q_1\cdot n_1)\times (\q_2\cdot n_2)}$, $\lambda(B)_{i+1,j+1} = \w(B\svtd{\q_1 i +1, \q_1}{\q_2 j + 1, \q_2})$, $0\leq i\leq n_1$, and $0\leq j\leq n_2$. Therefore, we have that $\w(B\svtd{\delta_1' i +1, \ell_1'}{\delta_2'j+1, \ell_2'}) = \w(\lambda(B)\svtd{\delta_1 i +1, \ell_1}{\delta_2j+1, \ell_2})$.
    For every $((\ell_1', \ell_2'), (\delta_1', \delta_2'))$-\Rcode\ $\mathbb{C}$, let $\mathbb{C}_{\lambda}$ be the code created by applying $\lambda$ on all the matrices of $\mathbb{C}$, i.e., $\mathbb{C}_\lambda = \{\lambda(B) : B\in \mathbb{C}\}$. 
    For every $B, D\in \mathbb{C}$, $\Rmat{\delta_1', \delta_2'}{\ell_1', \ell_2'}(B)\neq \Rmat{\delta_1', \delta_2'}{\ell_1', \ell_2'}(D)$, and thus, $\Rmat{\delta_1, \delta_2}{\ell_1, \ell_2}(\lambda(B))\neq\Rmat{\delta_1, \delta_2}{\ell_1, \ell_2}(\lambda(D))$. In conclusion, $\mathbb{C}_{\lambda}$ is an $((\ell_1, \ell_2), (\delta_1, \delta_2))_{\q}$-\Rcode. In addition, $\lambda(B)\neq \lambda(D)$, and thus, $|\mathbb{C}| = |\mathbb{C}_{\lambda}|$. Therefore,
    
    \vspace{-2ex}
    \begin{align}
        A(\q_1\cdot n_1, \q_2\cdot n_2, &(\ell_1', \ell_2'), (\delta_1', \delta_2')) \notag\\
        &\leq A_{\q}(n_1, n_2, (\ell_1, \ell_2), (\delta_1, \delta_2)).\label{eqn:2}
    \end{align}
    From (\ref{eqn:1}), (\ref{eqn:2}) we have that 
    \begin{align*}
        A(\q_1\cdot n_1, \q_2\cdot n_2, &(\ell_1', \ell_2'), (\delta_1', \delta_2')) \\
        &= A_{\q}(n_1, n_2, (\ell_1, \ell_2), (\delta_1, \delta_2)),
    \end{align*}
    and therefore, 
    \begin{align*}
        & \capacityq{\q}{(\ell_1, \ell_2), (\delta_1, \delta_2)} \\
        = & \frac{\log_\q A(\q_1\cdot n_1, \q_2\cdot n_2, (\ell_1', \ell_2'), (\delta_1', \delta_2'))}{n_1\cdot n_2} \\
        = & \frac{\q_1\cdot \q_2}{\log_2 \q}\frac{\log_2 A(\q_1\cdot n_1, \q_2\cdot n_2, (\ell_1', \ell_2'), (\delta_1', \delta_2'))}{\q_1\cdot n_1\cdot \q_2\cdot n_2} \\
        = & \frac{\q_1\cdot \q_2}{\log_2 \q}\capacity{(\ell_1', \ell_2'), (\delta_1', \delta_2')}.
    \end{align*}
\end{IEEEproof}
Theorem~\ref{theorem:binary-to-q-ary-two-d} implies that all the results in the two-dimensional binary model can be generalized to any $\q$-ary model as well.
Furthermore, we demonstrate in the next Theorems correlations between the capacity of the two-dimensional \Rchannel and that of the one-dimensional channel.
\begin{theorem}
\label{theorem:upper-lower-bounds}
If $\delta_1, \delta_2, \ell_1, \ell_2$ are integers such that $0<\delta_1 \leq \ell_1$, $0<\delta_2\leq \ell_2$
and $q$ is an alphabet size, then
$$
\capacityq{\q}{(\ell_1, \ell_2), (\delta_1, \delta_2)} \geq \capacityq{\q}{\delta_1\cdot \ell_2, \delta_1\cdot \delta_2}.
$$
\end{theorem}

\begin{IEEEproof}
Let $\delta_1, \delta_2, \ell_1, \ell_2$, be such integers, and let $n_1, n_2, t_1, t_2\in \mathbb{N}$ be positive integers
such that $n_i = t_i\delta_1 +\ell_i$. Let $\mathcal{C}$ be a $(\delta_1\cdot\ell_2, \delta_1\cdot\delta_2)$-\Rcode\
of length $n_2\cdot \delta_1$. For each $i$, $0\leq i\leq t_1$, let $\psii(i) = (i-1)\delta_1 + \ell_1+1$.
Let $\mathbb{C}\subseteq \Sigma_\q^{n_1\times n_2}$ be the code defined as follows.
For each sequence $V$ of length $t_1+1$ defined by $V = (\bvec{v}^0, \bvec{v}^1, \ldots, \bvec{v}^{t_1})$,
where $\bvec{v}^i$ is a word from $\mathcal{C}$, let $M(V)$ be an $n_1 \times n_2$ codeword in $\mathbb{C}$.
For every $0\leq k_1\leq t_1$, its sub-matrix $M(V)\svtd{\psii(k_1), \delta_1}{1, n_2}$ is
the column by column folding of the vector $\bvec{v}^{k_1} =(v'_1,v'_2,\ldots,v'_{n_2 \delta_1})$, i.e., 
    \begin{align*}
        M(V)\svtd{\psii(k_1), \delta_1}{1, n_2} \hspace{-0.5ex} = \hspace{-0.5ex}
        \begin{pmatrix}
            v'_1 & v_{\delta_1 + 1} & \cdots & v'_{(n_2-1)\cdot \delta_1 + 1} \\
            v'_2 & v_{\delta_1 + 2} & \cdots & v'_{(n_2-1)\cdot \delta_1 + 2} \\
            \vdots & \vdots & \ddots & \vdots \\
            v'_{\delta_1} & v'_{2\delta_1} & \cdots & v'_{n_2\cdot \delta_1} 
        \end{pmatrix}
    \end{align*}
and the sub-matrix $M(V)\sv{1, \ell_1-\delta_1}{1, n_2}$ is the all-zeros sub-matrix
(note that $\psii (0) = \ell_1 - \delta_1)$.
For two distinct words $V$, $U \subseteq \mathcal{C}^{t_1+1}$, we have that $M(V) \neq M(U)$ and therefore
$|\mathbb{C}| = |\mathcal{C}|^{t_1+1}$.
Let $V = (\bvec{v}^0, \bvec{v}^1, \ldots, \bvec{v}^{t_1})$, $U = (\bvec{u}^0, \bvec{u}^1, \ldots, \bvec{u}^{t_1})$
be two distinct words, where $\bvec{v}^i,\bvec{u}^i \in \mathcal{C}$.
Let $0\leq i\leq t_1$ be the smallest index such that $\bvec{v}^i \neq \bvec{u}^{i}$.
Since $\bvec{v}^i,\bvec{u}^{i}$ are both in $\mathcal{C}$, it follows that 
there exists an index $0\leq j\leq t_2$ such that
$$
\w((\bvec{v}^i)\sv{\delta_1\delta_2 j +1}{\delta_1\ell_2}) \neq \w((\bvec{u}^i)\sv{\delta_1\delta_2 j +1}{\delta_1\ell_2}),
$$
and therefore
$$
\w(M(V)\svtd{\psii(i),\delta_1}{ j\cdot \delta_2+1, \ell_2})\neq \w(M(U)\svtd{\psii(i),\delta_1}{ j\cdot \delta_2+1, \ell_2}).
$$
We also have that
\begin{small}
$$
M\hspace{-0.2ex}(V)\hspace{-0.2ex}\svtd{(i-1)\cdot \delta_1+1, \ell_1}{j\cdot \delta_2+1, \ell_2}\hspace{-0.6ex} = \hspace{-0.6ex}M\hspace{-0.2ex}(U)\hspace{-0.2ex}\svtd{(i-1)\cdot \delta_1+1, \ell_1-\delta_1}{j\cdot \delta_2+1, \ell_2}
$$
\end{small}
since $\bvec{v}^{i'} = \bvec{u}^{i'}$ for $i' < i$ by the assumption.
Therefore,
\begin{align*}
    \w(M(V)&\svtd{(i-1)\cdot \delta_1+1, \ell_1}{j\cdot \delta_2+1, \ell_2})\\
    &\neq \w(M(U)\svtd{(i-1)\cdot \delta_1+1, \ell_1}{j\cdot \delta_2+1, \ell_2})
\end{align*}
and hence
$$\Rmat{\delta_1, \delta_2}{\ell_1, \ell_2}(M(V)) \neq \Rmat{\delta_1, \delta_2}{\ell_1, \ell_2}(M(U)).
$$
Thus, $\mathbb{C}$ is an $((\ell_1, \ell_2), (\delta_1, \delta_2))$-\Rcode
and therefore,
$$
A(n_2\delta_1, \delta_1\ell_2, \delta_1\delta_2)^{(t_1+1)} \leq A(n_1, n_2, (\ell_1, \ell_2), (\delta_1, \delta_2)),
$$
which implies that
\begin{align*}
&\capacity{(\ell_1, \ell_2), (\delta_1, \delta_2)} \\ 
\geq & \lim_{n_1,n_2\rightarrow\infty} \frac{\log_\q A(n_2\delta_1, \delta_1\ell_2, \delta_1 \delta_2))^{t_1}}{n_1\cdot n_2} \\
= & \lim_{n_1,n_2\rightarrow\infty}\frac{\log_\q A(n_2\delta_1, \delta_1\ell_2, \delta_1 \delta_2))}{\delta_1 \cdot n_2}\\
= & \capacity{\delta_1\cdot \ell_2, \delta_1\cdot \delta_2}.
\end{align*}
\end{IEEEproof}

\begin{theorem}
For $\q\geq 2$, and $\delta_2\geq \ell_2$, we have that
$$
\capacityq{\q}{(\ell_1, \ell_2), (\delta_1, \delta_2)} = \frac{\ell_2}{\delta_2}\capacityq{\q}{\ell_2\cdot \ell_1, \ell_2\cdot \delta_1}.
$$
\end{theorem}
\begin{IEEEproof}
Since $\delta_2\geq \ell_2$, it follows that for every integers $n_1$, $n_2$, $t_1$, $t_2$,
such that $n_i = t_i\delta_i+\ell_i$ and a matrix $B\in \Sigma_\q^{n_1\times n_2}$, for each
$0\leq i< i' \leq t_2$, the sub-matrices $B\svtd{1, n_1}{\delta_2 \cdot i +1, \ell_2}$
and $B\svtd{1, n_1}{\delta_2 \cdot i' +1, \ell_2}$ do not overlap. 
Therefore, the values of each of the sub-matrices are independent, and hence the maximum number of possible
codewords in an $n_1\times n_2$ $((\ell_1, \ell_2), (\delta_1, \delta_2))$-read code, i.e.,
$A(n_1, n_2, (\ell_1, \ell_2), (\delta_1, \delta_2))$, is equal to the number of possible
$(t_2+1)$ $((\ell_1, \ell_2), (\delta_1, \delta_2))$-read matrices of size $n_1\times \ell_2$.
This is equal to the number of $(t_2+1)$ $(\ell_2\cdot \ell_1, \ell_2\cdot \delta_1)$-\rv s of size $n_1\cdot \ell_2$, i.e.,
$$
A(n_1, n_2, (\ell_1, \ell_2), (\delta_1, \delta_2)) = A(n_1\cdot \ell_2, \ell_2 \cdot \ell_1, \ell_1\cdot \delta_1)^{t_2}.
$$
Thus, 
\begin{align*}
        & \capacityq{\q}{(\ell_1, \ell_2), (\delta_1, \delta_2)} \\
        = & \lim_{n_1,n_2\rightarrow\infty} \frac{\log_\q  A(n_1\cdot \ell_2, \ell_2 \cdot \ell_1, \ell_1\cdot \delta_1)^{t_2}}{n_1\cdot n_2} \\
        = & \lim_{n_1,n_2\rightarrow\infty} \frac{t_2 \log_\q  A(n_1\cdot \ell_2, \ell_2 \cdot \ell_1, \ell_1\cdot \delta_1)}{n_1\cdot n_2} \\
        = & \lim_{n_1, n_2\rightarrow\infty} \frac{\log_\q  A(n_1\cdot \ell_2, \ell_2 \cdot \ell_1, \ell_2\cdot \delta_1)}{n_1\cdot \delta_2} \\
        = & \frac{\ell_2}{\delta_2}\capacityq{\q}{\ell_2\cdot \delta_1, \ell_2\cdot \delta_1}.
\end{align*}
\end{IEEEproof}
The following lemma is an immediate observation from the definitions.
\begin{lemma} \label{lemma:delta-1-ell-1-twodim}
    For integers $\q, \delta, \ell$, we have that,
    $$
    \capacityq{\q}{(\ell, 1), (\delta, 1)} = \capacityq{\q}{\ell, \delta}.
    $$
\end{lemma}

\begin{lemma} \label{lemma:delta-1-two-dim}
    For positive integers $\q, \delta, \ell_1, \ell_2$, we have that
    $$
    \capacityq{\q}{(\ell_1, \ell_2), (\delta, 1)} = \capacityq{\q}{\ell_1, \delta}.
    $$
\end{lemma}
\begin{IEEEproof}
    Let $n_1, n_2$ be integers, and let $\mathbb{C}$ be an $n_1\times n_2$ $((\ell_1, \ell_2), (\delta_1, 1))$-\Rcode.
    For every two distinct codewords $B, D\in \mathbb{C}$, we have that, $\Rmat{\ell_1, \ell_2}{\delta_1, 1}(B) \neq \Rmat{\ell_1, \ell_2}{\delta_1, 1}(D)$, and thus, $\Rmat{\ell_1, 1}{\delta_1, 1}(B) \neq \Rmat{\ell_1, 1}{\delta_1, 1}(D)$. In conclusion, $\mathbb{C}$ is an $((\ell_1, 1), (\delta_1, 1))$-\Rcode, and therefore,
    $$A(n_1, n_2, (\ell_1, \ell_2), (\delta_1, 1)) \leq A(n_1, n_2, (\ell_1, 1), (\delta_1, 1)),$$
    which implies that
    $$\capacity{(\ell_1, \ell_2), (\delta_1, 1)} \leq \capacity{(\ell_1, 1), (\delta_1, 1)}.$$ Thus, by Lemma~\ref{lemma:delta-1-ell-1-twodim} $$\capacity{(\ell_1, \ell_2), (\delta_1, 1)} \leq \capacity{\ell_1, \delta_1}.$$
    By Theorem~\ref{theorem:upper-lower-bounds}, we have that, 
    $$\capacity{(\ell_1,\ell_2), (\delta, 1)} \geq \capacity{\ell_1, \delta},$$
    and therefore, 
    $$\capacity{(\ell_1,\ell_2), (\delta, 1)} = \capacity{\ell_1, \delta}.$$
\end{IEEEproof}
\begin{theorem}
    For $\q\geq 2$, and $\delta_2$ divides $\ell_2$, we have that
        \[
            \capacityq{\q}{(\ell_1, \ell_2), (\delta_1, \delta_2)} = \capacityq{\q}{\delta_2\cdot \ell_1, \delta_2\cdot \delta_1}
        \]
\end{theorem}
\begin{IEEEproof}
    We start by showing the theorem for the binary case. Let $n_1, n_2, t_1, t_2\in \mathbb{N}$, such that $n_1 = t_1\delta_1+\ell_1$ and $n_2 = t_2\delta_2$.
    For every $n_1\times n_2$ $((\ell_1, \ell_2), (\delta_1, \delta_2))$-\Rcode\ $\mathbb{C}$, let $\mathbb{C}'\subseteq \Sigma_{\delta_2+1}^{n_1\times t_2}$ denote the set of all matrices $B'$ derived from the codewords $B\in \mathbb{C}$, such that for every $i, j$, such that $1\leq i \leq n_1$, $0\leq j \leq t_2-1$, $B'_{i, j+1} = \w(B\svtd{i, 1}{j\cdot \delta_2 +1, \delta_2})\in \Sigma_{\delta_2+1}$, were $B'$ is derived from the codeword $B\in \mathbb{C}$.
    We have that, $\Rmat{\delta_1, \delta_2}{\ell_1, \ell_2}(B) = \Rmat{\delta_1, 1}{\ell_1, \frac{\ell_2}{\delta_2}}(B')$. Therefore, $\mathbb{C}'$ is an $((\ell_1, \frac{\ell_2}{\delta_2}), (\delta_1, 1))_{\delta_2+1}$-\Rcode, and $|\mathbb{C}| = |\mathbb{C}'|$. Thus, 
    \begin{align*}
        A(n_1, \hspace{-0.3ex}n_2, \hspace{-0.2ex}(\ell_1\hspace{-0.2ex}, \ell_2), (\delta_1\hspace{-0.2ex}, \delta_2)) \hspace{-0.5ex}\leq\hspace{-0.3ex} A_{\delta_2+1}(n_1\hspace{-0.3ex}, \hspace{-0.3ex}t_2, \hspace{-0.2ex}(\ell_1, \frac{\ell_2}{\delta_2}), \hspace{-0.5ex}(\delta_1\hspace{-0.2ex}, \hspace{-0.3ex}1)). 
    \end{align*}
    For every $n_1\times t_2$ $((\ell_1, \frac{\ell_2}{\delta_2}), (\delta_1, 1))_{\delta_2+1}$-\Rcode\ $\mathbb{C}$, let $\mathbb{C}^*\subseteq \ABC^{n_1\times n_2}$ be a code derived from $\mathbb{C}$, such that for every $k_1, k_2$, such that $1\leq k_1\leq n_1$, $0\leq k_2\leq t_2-1$, $(B^*)\svtd{k_1, 1}{\delta_2k_2+1, \delta_2} = 0^{\delta_2-\alpha}1^{\alpha}$, where $\alpha = B_{k_1, k_2+1}$, and $B^*\in \mathbb{C}^*$ is derived from $B\in \mathbb{C}$.
    Thus, $\w((B^*)\svtd{k_1, 1}{\delta_2k_2+1, \delta_2}) =  \w(B_{k_1, k_2+1})$ and $\Rmat{\delta_1, \delta_2}{\ell_1, \ell_2}(B^*) = \Rmat{\delta_1, 1}{\ell_1, \frac{\ell_2}{\delta_2}}(B)$. Therefore, $\mathbb{C}^*$ is an $((\ell_1, \ell_2), (\delta_1, \delta_2))$-\Rcode, $|\mathbb{C}| = |\mathbb{C}^*|$ and
    \begin{align*}
        A(n_1\hspace{-0.2ex}, \hspace{-0.3ex} n_2, \hspace{-0.3ex} (\ell_1, \ell_2),\hspace{-0.3ex} (\delta_1, \delta_2)) \hspace{-0.5ex}\geq\hspace{-0.5ex} A_{\delta_2+1}(n_1, t_2, (\ell_1, \frac{\ell_2}{\delta_2}), \hspace{-0.3ex}(\delta_1\hspace{-0.2ex}, \hspace{-0.3ex} 1)) 
    \end{align*}
    Thus,
    \begin{align*}
        A(n_1\hspace{-0.2ex}, \hspace{-0.3ex} n_2, (\ell_1, \ell_2), \hspace{-0.3ex}(\delta_1\hspace{-0.2ex}, \delta_2))\hspace{-0.5ex} =\hspace{-0.5ex} A_{\delta_2+1}(n_1, t_2, (\ell_1\hspace{-0.2ex}, \frac{\ell_2}{\delta_2}),\hspace{-0.3ex} (\delta_1\hspace{-0.2ex}, \hspace{-0.3ex}1)) 
    \end{align*}
    and therefore, by lemma~\ref{lemma:delta-1-two-dim}, $\capacity{(\ell_1, \ell_2), (\delta_1, \delta_2)}$ is equals to
    \begin{align*}
        &\lim_{n_1,n_2\rightarrow\infty} \frac{\log_2A_\q(n_1, t_2, (\ell_1, \frac{\ell_2}{\delta_2}), (\delta_1, 1))}{n_1\cdot n_2} \\
        = & \frac{\log_2 \q}{\delta_2} \lim_{n_1,n_2\rightarrow\infty} \frac{\log_\q A_\q(n_1, t_2, (\ell_1, \frac{\ell_2}{\delta_2}), (\delta_1, 1))}{n_1\cdot t_2} \\
        = & \frac{\log_2 \q}{\delta_2} C_{\delta_2+1}((\ell_1, \frac{\ell_2}{\delta_2}), (\delta_1, 1)) \\
        = &\frac{\log_2 \q}{\delta_2} C_{\delta_2+1}(\ell_1, \delta_1) = \capacity{\ell_1, \delta_1}.
    \end{align*}
    Now, for every $\q\in \mathbb{N}$, by Theorem~\ref{theorem:binary-to-q-ary-two-d}
    \begin{align*}
        &\capacityq{\q}{(\ell_1, \ell_2), (\delta_1, \delta_2)} \\
        = &\frac{\q-1}{\log_2\q}\capacity{((\q-1)\ell_1,\ell_2), ((\q-1)\delta_1, \delta_2)} \\
        = & \frac{\q-1}{\log_2\q}\capacity{(\q-1)\ell_1, (\q-1)\delta_1}
        = \capacityq{\q}{\ell_1, \delta_1}.
    \end{align*}
\end{IEEEproof}

\begin{theorem}
\label{theorem:upper-lower-bounds-2}
For positive integers $\delta_1, \delta_2, \ell_1, \ell_2$, such that $0<\delta_i \leq \ell_i$, we have that
\begin{align*}
\capacityq{\q}{(\ell_1, \ell_2), (\delta_1, \delta_2)} \leq \capacityq{\q}{\ell_1, \delta_1}.
\end{align*}
\end{theorem}

\begin{IEEEproof}
Let $\delta_1, \delta_2, \ell_1, \ell_2$, such that $0<\delta_i \leq \ell_i$.
Let $n_1$, $n_2$, $t_1$, and $t_2$ be integers such that $n_i = t_i\delta_i+\ell_i$.
For every $n_1\times n_2$ $((\ell_1, \ell_2), (\delta_1, \delta_2))$-\Rcode\ $\mathbb{C}$ and
a codeword $B\in \mathbb{C}$, all sub-matrices of the form $B\svtd{k_1, \ell_1}{k_2, \ell_2}$,
where $k_1 = \delta_1 i +1$, $k_2 = \delta_2j + 1$, $0\leq i\leq t_1$ and $0\leq j\leq t_2$,
are also sub-matrices in $B\svtd{k_1, \ell_1}{k_2, \ell_2}$,
where $k_1 = \delta_1 i +1$, $k_2 = j + 1$, $0\leq i\leq t_1$ and $0\leq j\leq n_2-1$
and hence $\mathbb{C}$ is also an $((\ell_1, \ell_2), (\delta_1, 1))$-\Rcode. Therefore,
$$
\capacityq{\q}{(\ell_1, \ell_2), (\delta_1, \delta_2)} \leq \capacityq{\q}{(\ell_1, \ell_2), (\delta_1, 1)}.
$$
Thus, by Lemma~\ref{lemma:delta-1-two-dim} we have that
$$
\capacityq{\q}{(\ell_1, \ell_2), (\delta_1, \delta_2)} \leq \capacityq{\q}{\ell_1, \delta_1}.
$$
\end{IEEEproof}

\bibliographystyle{IEEEtran}
    \bibliography{IEEEabrv,bib}

\end{document}